\documentclass[fleqn,usenatbib]{mnras}

\usepackage{newtxtext,newtxmath}
\usepackage{siunitx}

\usepackage[T1]{fontenc}
\usepackage{ae,aecompl}

\DeclareRobustCommand{\VAN}[3]{#2}
\let\VANthebibliography\thebibliography
\def\thebibliography{\DeclareRobustCommand{\VAN}[3]{##3}\VANthebibliography}

\usepackage{graphicx}	% Including figure files
\usepackage{amsmath}	% Advanced maths commands
\usepackage[normalem]{ulem}

\usepackage{color} %costum addition 
\usepackage{hyperref} % custom addition
\usepackage{makecell} % costum addition
\newcommand*\pyorbit{\texttt{PyORBIT}}

\newcommand{\snr} {\mbox{S/N}}

\newcommand{\ie}{i.\,e.}

\newcommand{\teff}{$T_{{\rm eff}}$}
\newcommand{\kms}{\mbox{km\,s$^{-1}$}}
\newcommand{\ms}{\mbox{m s$^{-1}$}}
\newcommand{\gcm}{\mbox{g cm$^{-3}$}}

\newcommand{\vsini} {$v$\,sin\,$i$}

\newcommand{\vmicro} {\mbox{$\xi_{\rm t}$}}
\newcommand{\gfeh} {\mbox{$[{\rm Fe}/{\rm H}]$}}
\newcommand{\meh} {\mbox{$[{\rm M}/{\rm H}]$}}
\newcommand{\logg} {\mbox{log\,{\it g}}}

\newcommand{\mplanet}{\mbox{$M_{\rm p}$}}
\newcommand{\rplanet}{\mbox{$R_{\rm p}$}}
\newcommand{\rhoplanet}{\mbox{$\rho_{\rm p}$}}

\newcommand{\mearth}{\mbox{M$_\oplus$}}
\newcommand{\rearth}{\mbox{R$_\oplus$}}
\newcommand{\rhoearth}{\mbox{$\rho_\oplus$}}

\newcommand{\msun}{\mbox{M$_\odot$}}
\newcommand{\rsun}{\mbox{R$_\odot$}}
\newcommand{\mstar}{\mbox{$M_\star$}}
\newcommand{\rstar}{\mbox{$R_\star$}}

\newcommand{\rhostar}{\mbox{$\rho_\star$}}
\newcommand{\rhosun}{\mbox{$\rho_\odot$}}

\newcommand{\logRHK}{\mbox{$\log {\rm R}^{\prime}_{\rm HK}$}}

\newcommand{\starname}{\mbox{TOI-561}}
\newcommand{\ticname}{\mbox{TIC 377064495}}

\newcommand{\megno}{$\langle Y \rangle$}

\definecolor{darkcyan}{rgb}{0.0, 0.55, 0.55}
\definecolor{dartmouthgreen}{rgb}{0.05, 0.5, 0.06}
\definecolor{magenta}{rgb}{1., 0., 1.}

\title[TOI-561: an USP super-Earth and three mini-Neptunes]{An unusually low density ultra-short period super-Earth and three mini-Neptunes around the old star TOI-561}

\author[G. Lacedelli et al.]{
G. Lacedelli$^{1,2}$\thanks{E-mail: gaia.lacedelli@phd.unipd.it},
L. Malavolta$^{1,2}$,
L. Borsato$^{2}$,
G. Piotto$^{1,2}$,
D. Nardiello$^{3,2}$,
A. Mortier$^{4,5}$,
M. Stalport$^{6}$,
\newauthor
A. Collier Cameron$^{7}$,
E. Poretti$^{8,9}$,
L. A. Buchhave$^{10}$,
M. L\'opez-Morales$^{11}$,
V. Nascimbeni$^{2}$,
T. G. Wilson$^{7}$,
\newauthor
S. Udry$^{6}$,
D. W. Latham$^{11}$,
A. S. Bonomo$^{12}$,
M. Damasso$^{12}$,
X. Dumusque$^{6}$,
J. M. Jenkins$^{13}$,
C. Lovis$^{6}$,
\newauthor
K. Rice$^{14, 15}$,
D. Sasselov$^{11}$,
J. N. Winn$^{16}$,
G. Andreuzzi$^{8,17}$,
R. Cosentino$^{8}$,
D. Charbonneau$^{11}$,
\newauthor
L. Di Fabrizio$^{8}$,
A. F. Martinez Fiorenzano$^{8}$,
A. Ghedina$^{8}$,
A. Harutyunyan$^{8}$,
F. Lienhard$^{4}$,
G. Micela$^{19}$,
\newauthor
E. Molinari$^{18}$,
I. Pagano$^{20}$,
F. Pepe$^{6}$,
D. F. Phillips$^{11}$,
M. Pinamonti$^{12}$,
G. Ricker$^{21}$,
G. Scandariato$^{20}$,
\newauthor
A. Sozzetti$^{12}$,
C. A. Watson$^{22}$
\newauthor
\emph{\normalsize Affiliations are listed at the end of the paper}}
\date{Accepted 2020 November 27. Received 2020 November 26; in original form 2020 September 4.}

\pubyear{2020}

\begin{document}
\label{firstpage}
\pagerange{\pageref{firstpage}--\pageref{lastpage}}
\maketitle

% Abstract of the paper
\begin{abstract}
Based on HARPS-N radial velocities (RVs) and {\it TESS} photometry, we present a full characterisation 
of the planetary system orbiting the 
late G dwarf
\starname. 
After the identification of three transiting candidates by {\it TESS}, we discovered two additional external planets from RV analysis. 
RVs cannot confirm the outer {\it TESS} transiting candidate, which would also make the system dynamically unstable. We demonstrate that the two transits initially associated with this candidate are instead due to single transits of 
the two planets discovered using RVs.
The four planets orbiting \starname\ include an ultra-short period (USP) super-Earth (\starname\ b) with period $P_{\rm b} = 0.45$~d, mass $M_{\rm b} =1.59 \pm 0.36$~\mearth\ and radius $R_{\rm b}=1.42 \pm 0.07$~\rearth, and three mini-Neptunes:
\starname\ c, with  $P_{\rm c} = 10.78$~d, $M_{\rm c} =5.40 \pm 0.98$~\mearth, $R_{\rm c}= 2.88 \pm 0.09$~\rearth;
\starname\ d, with $P_{\rm d} = 25.6$~d, $M_{\rm d} = 11.9 \pm 1.3$~\mearth, $R_{\rm d} = 2.53 \pm 0.13$~\rearth;
and \starname\ e, with $P_{\rm e} = 77.2$~d, $M_{\rm e} = 16.0 \pm 2.3$~\mearth, $R_{\rm e} = 2.67 \pm 0.11$~\rearth.
Having a density of $3.0 \pm 0.8$~\gcm, TOI-561 b is the lowest density USP planet known to date. 
Our N-body simulations confirm the stability of the system and predict a strong, anti-correlated, long-term transit time variation signal 
between planets d and e. 
The unusual density of the inner super-Earth and the 
dynamical interactions between the outer planets make \starname\ an interesting follow-up target.
\end{abstract}

% Select between one and six entries from the list of approved keywords.
% Don't make up new ones.

%%%%%%%%%%%%%%%%%%%%%%%%%%%%%%%%%%%%%%%%%%%%%%%%%%

%%%%%%%%%%%%%%%%% BODY OF PAPER %%%%%%%%%%%%%%%%%%

\section{Introduction}\label{sec:introduction}

The {\it Transiting Exoplanet Survey Satellite} \cite[{\it TESS},][]{ricker2014} is a NASA all-sky survey designed to search for transiting planets around bright and nearby stars, and particularly
targeting stars that could reveal planets with radii smaller than Neptune.
Since the beginning of its observations in 2018, {\it TESS} has discovered more than $66$ exoplanets, including about a dozen multi-planet systems \cite[e.g.][]{Dragomir_2019, Dumusque2019, gunter2019}. 
Multi-planet systems, orbiting the same star and having formed from the same protoplanetary disc,
offer a unique opportunity for comparative planetology. 
They allow for investigations of the formation and evolution processes, i.e. through studies of 
relative planet sizes and orbital separations, 
orbital inclinations relative to the star's rotation axis, mutual inclination of the orbits, etc. 
In order to obtain a complete characterisation of a system, knowledge of the orbital architecture and the bulk composition of the planets are essential.  
To obtain such information, transit photometry needs to be combined with additional techniques that allow for the determination of the planetary masses, i.e. radial velocity (RV) follow-up or transit time variation (TTV) analysis.
Up to now, the large majority of known planetary systems have been discovered by the \emph{Kepler} space telescope \citep{Borucki2010}, which has led to an unprecedented knowledge of the ensemble properties of multiple systems \citep[e.g.][]{latham11, millholland2017, weiss2018}, their occurrence rate \citep[e.g.][]{Fressin_2013}, and their dynamical configurations \citep[e.g.][]{Lissauer_2011, Fabrycky_2014}. However, many of the \emph{Kepler} targets are too faint for RV follow-up, so most of the planets do not have a mass measurement, preventing a comprehensive understanding 
of their properties, and of the planetary system.
Thanks to the {\it TESS} satellite, which targets brighter stars, an increasing number of
candidates suitable for spectroscopic follow-up campaigns are being discovered. 
These new objects will increase the number of well characterised systems, and will provide a valuable observational counterpart to the theoretical studies on the formation and evolution processes 
of planetary systems
\citep[e.g.][]{morbidelli2012, Raymond_2014, helled2014, Baruteau2014prpl.conf..667B, Baruteau2016SSRv..205...77B, Davies2014prpl.conf..787D}.\\
In this paper, we combine {\it TESS} photometry (Section~\ref{sec:tess_photometry}) and high precision RVs gathered with 
the HARPS-N spectrograph (Section~\ref{sec:harpsn_RV})
to characterise the multi-planet system orbiting the star \starname.
The {\it TESS} pipeline identified three candidate planetary signals, namely an ultra-short period (USP) candidate ($P \sim 0.45$ days), and two additional candidates with periods of $\sim 10.8$ and $\sim 16.4$ days.
We determined the stellar properties (Section~\ref{sec:stellar_params}) using three independent methods. 
Based on our activity analysis, we 
concluded
that \starname\ is an
old, quiet star,
and therefore quite appropriate for the study of a complex
planetary system.
After assessing the planetary nature of the transit-like features (Section~\ref{sec:false_positives}), we performed a series of analysis~--~with the tools described in Section~\ref{sec:data_analysis_tools}~--~to determine the actual system configuration (Section~\ref{sec:system_architecture}).
We further address the robustness of our final solution based on a comparison with other possible models (Section~\ref{sec:model_comparison}). 
We finally compare the resulting planetary densities with the distribution of known planets in the mass-radius diagram and we predict the expected TTV signal for the planets in the system (Section~\ref{sec:discussion_conclusions}).

\section{Observations}\label{sec:observations}

\subsection{{\it TESS} photometry}\label{sec:tess_photometry}
\starname\ was observed by {\it TESS} in two-minute cadence mode
during observations of 
sector $8$, 
between 2 February and 27 February 2019. 
The astrometric and photometric 
parameters
of the star are listed in Table~\ref{table:star_astrometric_params}.
Considering the download time, and the loss of $3.26$ days of data due to an interruption in communications
between the instrument and 
the
spacecraft that occurred during sector $8$\footnote{See {\it TESS} Data Release Notes: Sector $8$, DR10 (\url{https://archive.stsci.edu/tess/tess_drn.html}).},
a total of $20.22$ days of science data were collected. 
The photometric observations for \starname\
were reduced by the Science Processing Operations Center (SPOC) 
pipeline \citep{jenkins2016,jenkins2020},
which detected three candidate planetary 
signals,
with periods of $10.8$ days (TOI-561.01),
$0.4$ days (TOI-561.02), and $16.4$ days (TOI-561.03),
respectively.
The pipeline identified $55$ transits of TOI-561.02, two 
transits
of TOI-561.01, and two 
transits
of TOI-561.03,
with depths of $290$, $1207$, and $923$ ppm and signal-to-noise-ratios (\snr) of $10.0$, $9.8$ and $9.2$, respectively. 
For our photometric analysis, 
we used the light curve based on the Pre-search Data Conditioning Simple Aperture Photometry
\citep[PDCSAP, ][]{smith2012, Stumpe2012, Stumpe2014}.
We downloaded the two-minute cadence PDCSAP light 
curve from the
Mikulski Archive for Space Telescopes (MAST)\footnote{
\url{https://mast.stsci.edu/portal/Mashup/Clients/Mast/Portal.html}},
and removed all the observations
encoded as \emph{NaN} or flagged as bad-quality (\texttt{DQUALITY>0}) points by the SPOC pipeline\footnote{ \url{https://archive.stsci.edu/missions/tess/doc/EXP-TESS-ARC-ICD-TM-0014.pdf}}.
We performed outliers rejection by doing a cut at $3 \sigma$ for positive outliers
and $5 \sigma$ (\ie\ larger than the deepest transit) for negative outliers. 
We removed the low frequency trends in the light curve using the biweight time-windowed slider implemented in the \texttt{wotan} package \citep{hippke2019}, with a window of $1.5$ days,
and masking the in-transit points to avoid modifications of the transit shape.
In order to obtain an independent confirmation of the signals detected in the {\it TESS} light curve, 
we performed an iterative transit search on the detrended light curve using the Transit Least Squares (TLS) algorithm \citep{hippke2019_TLS}. The first three significant identified signals nicely matched the {\it TESS} suggested periods ($P_\mathrm{TLS} = 10.78$~d, $0.44$~d, $16.28$~d).\\
In addition, we also extracted the $30$-minutes cadence light curve from the {\it TESS}
Full-Frame Images (FFIs) using the \texttt{PATHOS} pipeline \citep{nardiello2019},
in order to obtain an independent confirmation of the detected signals 
(Section~\ref{sec:false_positives}).

\begin{table}
\caption{Astrometric and photometric 
parameters
of \starname }
\label{table:star_astrometric_params}
\centering
\begin{tabular}{lcc}
\hline\hline
Property & Value & Source \\
\hline
\multicolumn{3}{c}{\emph{Other target identifiers}}\\
\hline
TIC & 377064495 & A\\
{\it Gaia} DR2 & 3850421005290172416 & B\\
2MASS & J09524454+0612589 & C\\
\hline
\multicolumn{3}{c}{\emph{Astrometric 
parameters
}}\\
\hline
RA  (J2015.5; h:m:s)&  09:52:44.44 & B\\
Dec (J2015.5; d:m:s)& 06:12:57.97 & B\\
$\mu_{\alpha}$ (mas yr$^{-1}$) & $-108.432 \pm 0.088$ & B\\
$\mu_{\delta}$ (mas yr$^{-1}$) & $-61.511 \pm 0.094$ & B\\
Systemic velocity (\kms ) & $79.54 \pm 0.56$ & B \\
Parallax$^a$ (mas) & $11.6768 \pm 0.0672 $& B\\
Distance (pc) & $85.80_{-0.49}^{+0.50}$& D \\
\hline
\multicolumn{3}{c}{\emph{Photometric 
parameters
}}\\
\hline
{\it TESS} (mag) & $9.527 \pm 0.006$& A\\
{\it Gaia} (mag) & $10.0128 \pm 0.0003$ & B\\
{\it V } (mag) & $10.252 \pm 0.006$ & A\\
{\it B }(mag) & $10.965 \pm	0.082$ & A\\
{\it J }(mag) & $8.879  \pm	0.020$ & C\\
{\it H }(mag) & $8.504  \pm	0.055$ & C\\
{\it K }(mag) & $8.394  \pm	0.019$ & C\\
{\it W1} (mag) & $8.337  \pm	0.023$ & E\\
{\it W2} (mag) & $8.396  \pm	0.020$ & E\\
{\it W3} (mag) & $8.375  \pm	0.023$ & E\\
{\it W4} (mag) & $7.971  \pm	0.260$ & E\\
\hline
Spectral type & G9V & F\\
\hline
\multicolumn{3}{l}{A) {\it TESS} Input Catalogue Version 8 (TICv8, \citealt{Stassun2018}).}\\
\multicolumn{3}{l}{B) {\it Gaia} DR2 \citep{GaiaColl2018}.}\\
\multicolumn{3}{l}{C) Two Micron All Sky Survey (2MASS, \citealt{Cutri2003}).}\\
\multicolumn{3}{l}{D) \cite{bailer-jones2018}.}\\
\multicolumn{3}{l}{E) {\it Wide-field Infrared Survey Explorer} \citep[{\it WISE};][]{Wright2010}.}\\
\multicolumn{3}{l}{\makecell[l]{ F) Based on \citet{pecaut2013}, assuming {\it Gaia} DR2,\\ Johnson,  2MASS and {\it WISE} color indexes.}}\\
\multicolumn{3}{l}{\makecell[l]{$^a$ {\it Gaia} DR2 parallax is corrected by $+ 50 \pm 7$~$\mu$as (with the error added \\ in quadrature) as suggested by \cite{Khan2019}.}}\\
\end{tabular}
\end{table}

\subsection{HARPS-N spectroscopy}\label{sec:harpsn_RV}
We collected $82$\footnote{
$62$ 
spectra
were collected within the Guaranteed Time Observations (GTO) time \citep{Pepe2013}, 
while the remaining $20$ 
spectra
were collected within the A40\_TAC23 program.} 
spectra using HARPS-N at the Telescopio
Nazionale Galileo (TNG), in La Palma \citep{Cosentino2012, Cosentino2014},
with the goal of precisely determining the masses of the three candidate planets and to search for additional 
planets. 
The observations started on 
November 17, 2019
and ended on 
June 13, 2020,
with an interruption between the end of March and the end of April due to the shut down of the TNG because of Covid-$19$.
In order to precisely characterise the signal of the 
USP
candidate, 
we collected $6$ points per night on 
February 4 and February 6, 2020,
thus covering the whole phase curve of the planet, and two points per night (when weather allowed) during the period of maximum visibility of the target (February-March 2020). 
The exposure time was set to $1800$ seconds, which resulted in a \snr~at 550~nm of~$77 \pm 20$ (median $\pm$ standard deviation) and a measurement uncertainty of $1.2 \pm 0.6$~\ms.
We reduced the data using the standard HARPS-N Data Reduction Software (DRS) using a {\tt G2} flux template (the closest match to the spectral type of our target) to correct for variations in the flux distribution as a function of the wavelength, and a {\tt G2} binary mask to compute the cross-correlation function
\citep[CCF,][]{Baranne1996, Pepe2002}. 
All the observations were gathered with the second fibre of HARPS-N 
illuminated by the Fabry-Perot calibration lamp to correct for the instrumental RV drift, except for the night of 
May 31, 2020. 
This observation setting prevented us from using the second fibre to correct for Moon contamination. 
However, we note that the difference between the systemic velocity of the star and the Moon is always greater than $15$~\kms , therefore preventing any contamination of the stellar CCF (as empirically found by \citealt{Malavolta2017a} and subsequently demonstrated through simulations by \citealt{Roy2020}),
as the average full width at half maximum (FWHM) of the CCF for TOI-561 is $6.380 \pm 0.004$~\kms. \\
The RV data with their $1\sigma$ uncertainties 
and the associated activity indices (see Section~\ref{sec:stellar_activity} for more details) are listed in Table~\ref{table:RV_table}. 
Before proceeding with the analysis, we removed from the total dataset $5$ RV measurements, with associated errors greater than $2.5$~\ms\ from spectra with \snr~$< 35$, that may affect the accuracy of our results. The detailed procedure performed to identify these points is described in Appendix~\ref{sec:appendix_removal}.

\begin{table*}
  \caption{HARPS-N Radial Velocity Measurements.}
\label{table:RV_table}      % is used to refer this table in the text
\centering                                      % used for centering table
\begin{tabular}{c c c c c c c c c}          % centered columns 
\hline\hline                        % inserts double horizontal lines
  BJD$_{\rm TDB}$ & RV & $\sigma _{\text RV}$ & BIS & FWHM  & $V_\mathrm{asy}$ & $\Delta V$ & \logRHK\ & H$\alpha$\\
 $($d$)$  & (\ms ) & (\ms ) & (\ms ) & (\kms ) & & (\kms) & & (dex)\\
\hline                                 % inserts single horizontal
2458804.70779  &  79700.63  &  1.27  &  -39.98  &  6.379  & 0.048 & -0.039 & -5.005 & 0.203\\
2458805.77551  &  79703.74  &  0.97  &  -36.25  &  6.380  & 0.049 & -0.036 & -4.984 & 0.200\\
2458806.76768  &  79701.71  &  1.05  &  -31.81  &  6.378  & 0.045 & -0.033 & -5.000 & 0.200\\
... & ... & ... & ... & ... & ... & ... & ... & ...\\
\hline
\multicolumn{9}{l}{This table is available in its entirety in machine-readable form.}
\end{tabular}
\end{table*}

\section{Stellar parameters}\label{sec:stellar_params}
\subsection{Photospheric parameters}\label{sec:stellar_photospheric_params}
We derived the photospheric stellar parameters using three different techniques: the curve-of-growth approach, spectral synthesis match, and empirical calibration.\\
The first method minimizes the trend of iron abundances (obtained from the equivalent width, EW, of each line) with respect to excitation potential and reduced EW respectively, to obtain the effective temperature and the microturbulent velocity, \vmicro. The gravity \logg\ is obtained by imposing the same average abundance from neutral and ionised iron lines. 
We obtained the EW measurements using {\tt  ARESv2}\footnote{Available at \url{http://www.astro.up.pt/~sousasag/ares/}} \citep{Sousa2015}.
We used the local thermodynamic equilibrium (LTE) code {\tt  MOOG}\footnote{Available at \url{http://www.as.utexas.edu/~chris/moog.html}} \citep{Sneden1973} for the line analysis, together with the {\tt ATLAS9} grid of stellar model atmosphere from \cite{Castelli2003}. The whole procedure is described in more detail in \citet{sousa2014}.
We performed the analysis on a co-added spectrum (\snr $> 600$),
and after applying the gravity correction from \cite{Mortier2014} and adding systematic errors in quadrature \citep{Sousa2011}, we obtained  \teff~$ = 5346 \pm 69 $~K, \logg~$ = 4.60 \pm 0.12$, \gfeh~$ = -0.40 \pm 0.05$ and \vmicro~$= 0.78 \pm 0.08$~\kms.\\
The spectral synthesis match was performed using the Stellar Parameters Classification tool ({\tt SPC}, \citealt{Buchhave2012, Buchhave2014}). 
It determines effective temperature, surface gravity, metallicity and line broadening by performing a cross-correlation of the observed spectra with a library of synthetic spectra, and interpolating the correlation peaks to determine the best-matching parameters. 
For technical reasons, we ran the {\tt SPC} on the $62$ GTO spectra only\footnote{{\tt SPC} runs on a server with access to GTO data only, and the required technical effort to enable the use of A40\_TAC23 data, complicated by the global Covid-19 sanitary emergency, was not justified by the negligible scientific gain.}: the \snr\ is so high that 
the spectra are anyway dominated
by systematic errors, and including the A40TAC\_23 spectra would not change the results.
We averaged the values measured for each exposure, and we obtained \teff~$ = 5389 \pm 50 $~K, \logg~$ = 4.49 \pm 0.10$, \meh~$= - 0.36 \pm 0.08$ and  \vsini~$< 2$~\kms.\\
We finally used \texttt{CCFpams}\footnote{Available at \url{https://github.com/LucaMalavolta/CCFpams}}, a method based on
the empirical calibration of temperature, metallicity and gravity
on several CCFs obtained with subsets of stellar lines with different sensitivity to temperature \citep{Malavolta2017b}.
We obtained \teff~$ = 5293 \pm 70 $~K, \logg~$ = 4.50 \pm 0.15 $
and \gfeh~$ = -0.40 \pm 0.05$, after applying the same gravity and systematic corrections as for the EW analysis.\\
We list the final spectroscopic adopted values, \ie, the weighted averages of the three methods, in Table~\ref{table:star_derived_params}. \par
From the co-added HARPS-N spectrum, we also derived the chemical abundances for several refractory elements (Na, Mg, Si, Ca, Ti, Cr, Ni). 
We used the ARES+MOOG method assuming LTE, as described earlier.
The reference for solar values was taken from \citet{asplund2009}, and all values in Table~\ref{table:star_derived_params} are given relative to the Sun. 
Details on the method and line lists are described in \citet{Adibekyan2012} and \citet{Mortier2013c}. 
This analysis shows that this iron-poor star is alpha-enhanced. 
Using the average abundances of magnesium, silicon, and titanium to represent the alpha-elements and the iron abundance from the ARES+MOOG method (for consistency), we find that $[\alpha/$Fe$] = 0.23$.

\subsection{Mass, radius, and density of the star}\label{sec:stellar_mass_radius}

For each set of photospheric parameters, we determined the stellar mass and radius using {\tt isochrones} \citep{Morton2015}, with posterior sampling performed by {\tt MultiNest} \citep{Feroz2008,Feroz2009,Feroz2019}. We provided as input the parallax of the target from the {\it Gaia} DR2 catalogue, after adding an offset of $+ 50 \pm 7$~$\mu$as (with the error added in quadrature to the parallax error) as suggested by \cite{Khan2019}, plus the photometry from the TICv8, 2MASS and {\it WISE} (Table~\ref{table:star_astrometric_params}). We used two evolutionary models, the MESA Isochrones \& Stellar Tracks (MIST, \citealt{Dotter2016,Choi2016,Paxton2011}) and the Dartmouth Stellar Evolution Database \citep{Dotter2008}.
For all methods, we assumed $\sigma_{T_{{\rm eff}}} = 70$~K, $\sigma_{\text{log\,{\it g}}} = 0.12$, $\sigma_{[{\rm Fe}/{\rm H}]} = 0.05$ (except for \texttt{SPC}, where we kept the original error of $0.08$) as a good estimate of the systematic errors regardless of the internal error estimates, to avoid favouring one technique over the others when deriving the stellar mass and radius. We also imposed an upper limit on the age of $13.8$~Gyr, \ie\ the age of the Universe \citep{Planck2018}.
From the mean and standard deviation of all the posterior samplings we obtained  \mstar~$=0.785 \pm 0.018$~\msun\ and \rstar~$= 0.849 \pm 0.007$~\rsun. We derived the stellar density \rhostar~$ = 1.285 \pm 0.040$~\rhosun\ (\rhostar~$= 1.809 \pm 0.056$~\gcm) directly from the posterior distributions of \mstar\ and \rstar.\\
We summarise the derived astrophysical parameters of the star in Table~\ref{table:star_derived_params}, which also reports temperature, gravity and metallicity obtained from the posteriors distributions resulting from the \texttt{isochrone} fit.
A lower limit on the age of $\sim 10$~Gyr is obtained considering the $15.86$-th percentile of the distribution of the combined posteriors, as for the other parameters.
We note however that an isochrone fit performed through {\tt EXOFASTv2} \citep{eastman2019}, assuming the photometric parameters in Table~\ref{table:star_astrometric_params} and the  spectroscopic parameters in Table~\ref{table:star_derived_params}, using only the MIST evolutionary set, returned a lower limit on the age of $5$~Gyr, while all the other parameters were consistent with the results quoted in Table~\ref{table:star_derived_params}. Thus, we decided to assume $5$~Gyr as a conservative lower limit for the age of the system.
The old stellar age and the sub-solar metallicity suggest that \starname\ may belong to an old Galactic population, an hypothesis that is also supported by our kinematic analysis. 
In fact, we derived the Galactic space velocities using the astrometric properties reported in Table~\ref{table:star_astrometric_params}. For the calculations we used the \texttt{astropy} package, and we assumed the {\it Gaia} DR2 radial velocity value of $79.54$~\kms, obtaining the heliocentric velocity components $(U, V, W) = (-60.0, -70.8, 16.7)$~\kms, in the directions of the Galactic center, Galactic rotation, and north Galactic pole, respectively.
The derived $UVW$ velocities point toward a thick-disk star, as confirmed by the probability membership derived following \citet{bensby2014}, that implies a $\sim 70$\% probability that the star belongs to the thick disc, a $\sim 29$\% probability of being a thin-disc star and a $\sim 0.0004$\% probability of belonging to the halo.

\begin{table}
  \caption{Derived astrophysical stellar parameters.}
\label{table:star_derived_params}      
\centering                                     
\begin{tabular}{l c c}          
\hline\hline                        
Parameter & Value & Unit \\
\hline                               
\noalign{\smallskip}
\teff$_{\rm spec}^a$ & $5372 \pm 70$ & K \\
\logg$_{\rm spec}^a$ & $4.50 \pm 0.12$& - \\
\gfeh$_{\rm spec}^a$ & $-0.40 \pm 0.05$ & -\\
\teff$^b$ & $ 5455_{-47}^{+65} $ & K \\ 
\logg$^b$  & $ 4.47 \pm 0.01$ & -  \\ 
\gfeh$^b$  & $ -0.33_{-0.05}^{+0.10}$ & - \\
\rstar\  & $ 0.849 \pm 0.007 $ & \rsun \\
\mstar\ & $ 0.785 \pm 0.018 $ & \msun \\
\rhostar & $ 1.285 \pm 0.040$ & \rhosun  \\
\rhostar & $1.809 \pm 0.056$ & \gcm\\
$A_V$  & $0.12_{-0.06}^{+0.08}$ & mag \\ 
\vsini & $< 2$ & \kms\\
age$^{c}$  & $ > 5 $ & Gyr \\ 
\logRHK & $-5.003 \pm 0.012$ & - \\
\hline
$[$Na/H$]$ & $-0.28 \pm 0.06$ & - \\
$[$Mg/H$]$ & $-0.17 \pm 0.05$ & - \\
$[$Si/H$]$ & $-0.22 \pm 0.05$ & - \\
$[$Ca/H$]$ & $-0.27 \pm 0.06$ & - \\
$[$Ti/H$]$ & $-0.12 \pm 0.03$ & - \\
$[$Cr/H$]$ & $-0.33 \pm 0.08$ & - \\
$[$Ni/H$]$ & $-0.37 \pm 0.04$ & - \\
\hline
\multicolumn{3}{l}{\makecell[l] {$^a$ Weighted average of the three spectroscopic \\methods.}}\\
\multicolumn{3}{l}{$^b$ Value inferred from the isochrone fit.}\\
\multicolumn{3}{l}{$^c$ Conservative lower limit.}
\end{tabular}
\end{table}

\subsection{Stellar activity}\label{sec:stellar_activity}
The low value of the \logRHK\ index ($-5.003 \pm 0.012$), derived using the calibration by \cite{Lovis2011} and assuming $B-V=0.71$,
indicates that \starname\ is a relatively quiet star. Given its distance of $\simeq 86$~pc, the lack of interstellar absorption near the Na D doublet in the HARPS-N co-added spectrum, and the total extinction in the {\it V} band from the isochrone fit
($0.1$~mag), 
we do not expect any significant effect of the interstellar medium on the \logRHK\ index \citep{Fossati2017}. Nevertheless, it is important to check
whether
the star is showing any sign of activity in all the activity diagnostics at our disposal.
In addition to the \logRHK index, FWHM, and bisector span (BIS) computed by the HARPS-N DRS, we included in our analysis the $V_\mathrm{asy}$ \citep{Figueira2013} and $\Delta V$ \citep{Nardetto2006} asymmetry indicators, as implemented by \citet{lanza2018}, and the chromospheric activity indicator H$\alpha$ \citep{GomesDaSilva2011}.\par
\begin{figure}
\centering
  \includegraphics[width=\hsize]{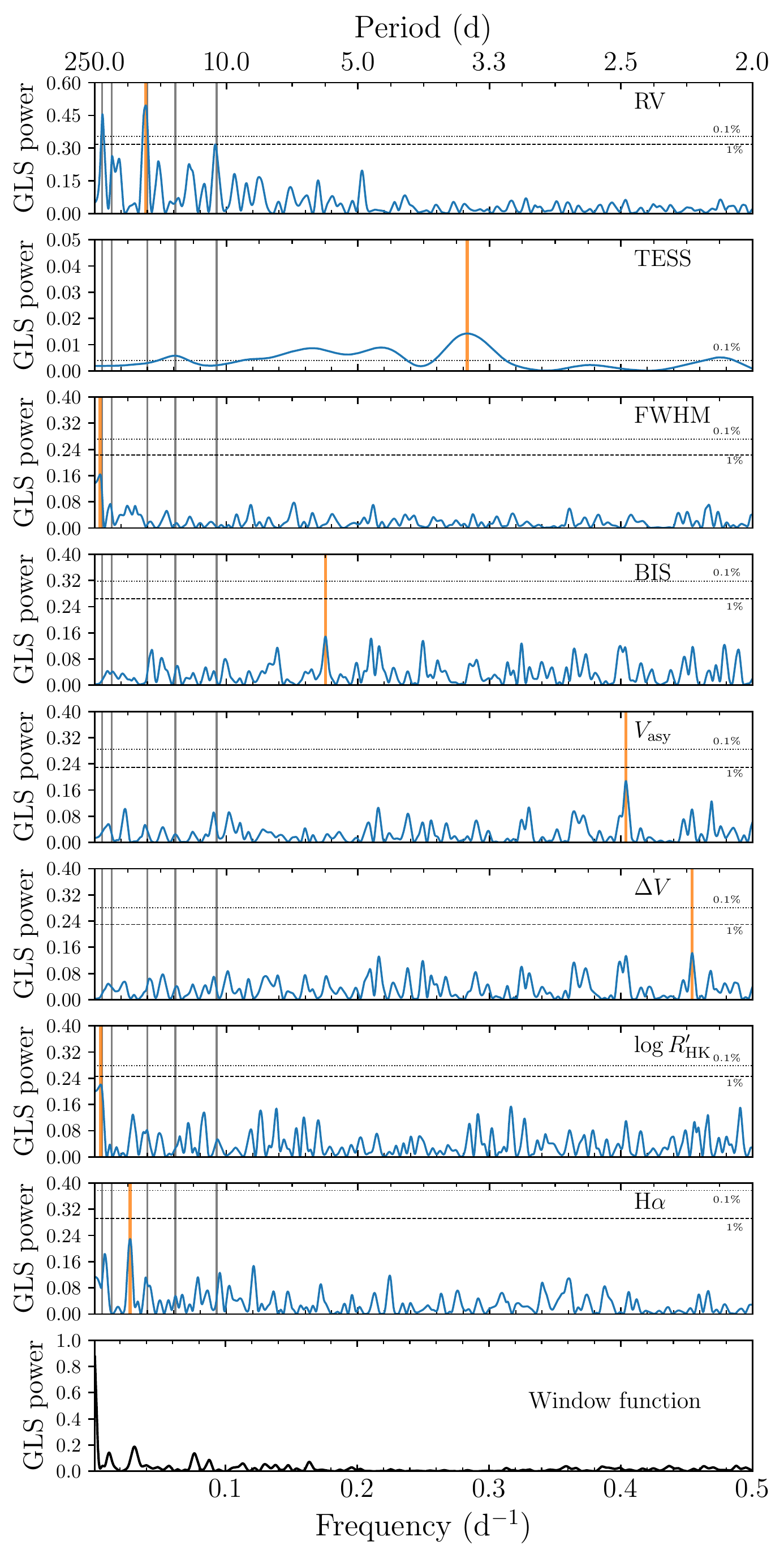}
  \caption{GLS periodogram of the RVs, the {\it TESS} photometry (PDCSAP) and the spectroscopic activity indexes under analysis. 
  The main peak of each periodogram is highlighted with an orange vertical line.
  The grey vertical lines represent the signals corresponding to the transit-like signals with periods $10.8$ and $16.3$ days, 
  and the additional signals identified in the RVs 
  (Section~\ref{sec:system_architecture})
  at $\simeq 25$, $\simeq 78$ and $\simeq 180$ days. 
  The dashed and dotted horizontal lines show the $1\%$ and $0.1\%$ FAP levels, respectively. 
  The {\it TESS} periodogram shows a series of peaks below $10$ days, unlikely to be associated with stellar activity given the old age of the star.
  The FWHM and the \logRHK\ periodograms have the main peak at $244$ and $220$ days, respectively,
  so there is no correspondence with the $180$ days signal. 
  Moreover, both of them are below the $1\%$ FAP. 
  The bottom panel shows the window function of the data.
    }
    \label{fig:activity_periodograms}
\end{figure}

 
The Generalized Lomb-Scargle (GLS, \citealt{Zechmeister2009}) periodograms of the above-mentioned indexes, computed within the frequency range $0.0005$--$0.5$~d$^{-1}$, \ie, $2$--$2000$ days, are shown in Figure~\ref{fig:activity_periodograms}, together with the periodograms of the RVs and {\it TESS} photometry. 
For each periodogram, we also report the power threshold corresponding to a False Alarm Probability (FAP) of $1$\% and $0.$1\%, computed with a bootstrap approach. 
The periodogram of the RVs reveals the presence of significant peaks at $\simeq 25$ days, $\simeq 180$ days, $\simeq 10$ days (corresponding to one of the transiting planet candidates), and $\simeq 78$ days, ordered decreasingly according to their power. 
None of these peaks has a counterpart in the activity diagnostics here considered, as no signals with a FAP lower than $2.4$\% can be identified, strongly supporting that the signals in the RVs are not
related
to stellar activity.
We note that the GLS periodogram of the {\it TESS} light curve identified a periodicity around $3.5$ days with an amplitude of $0.13$~ppt and a power of $0.014$, that is, above the $0.1$\% FAP threshold. However, it is unlikely that such variability is associated with stellar activity, since a rotational period of just a few days would be extremely atypical for a star older than $1$~Gyr (e.g. \citealt{douglas2019}), and in contrast with the lack of any signal in all the other above-mentioned activity indicators. Indeed, the rotational period estimated from the \logRHK\ using the calibrations of \citet{noyes1984} and \citet{mamajek2008} supports this assertion, indicating a value around $33$~d. We note that this value of the rotational period should be considered as a rough estimate, also because these calibrations are not well tested for old and alpha-enhanced stars like TOI-561. Further evidence against a $\sim 3.5$~d rotational period is provided by the low value of the \vsini\ ($< 2$~\kms), that suggests a rotational period $> 21.5$~d, assuming the stellar radius listed in Table~\ref{table:star_derived_params} and an inclination of $90$\degr. 
In any case, we verified with a periodogram analysis that our light curve flattening procedure correctly removed the here identified signal at $3.5$ days. \par
In addition, we performed an auto correlation analysis, following the prescription
by
\cite{McQuillan2013}, on the {\it TESS} light curve (with the transits filtered out), and the ASAS-SN {\it V} and {\it g} photometry \citep{Shappee2014,Kochanek2017}, after applying a $5$-$\sigma$ filtering,
but no significant periodicity could be identified.
A periodogram analysis of the ASAS-SN light curves in each band, either by taking the full dataset or by analysing each observing season individually, confirmed these results. \par
In conclusion, if any activity is present, its signature must be below $0.8$~ppt in the short period (rotationally-induced activity, $<$ 30 days), and $20$~ppt in the long term period (magnetic cycles, $>$ 100 days), from the RMS of {\it TESS} and ASAS-SN photometry respectively.
Incidentally, the former is close to the photometric variations of the Sun during the minimum at the end of Solar Cycle $25$, when the Sun also reached a \logRHK\ very close to the one measured for \starname\ \citep{CollierCameron2019,Milbourne2019}. By comparing our target to the Sun, and in general by taking into account the results of \cite{Isaacson2010}, 
it is expected that the contribution to the RVs due to the magnetic activity of our star is likely below $1$-$2$~\ms .
Since this value is quite close to the median internal error of our RVs,
no hint of the rotational period is provided by either the photometry or the spectroscopic activity diagnostics,
and the low activity level is consistent with our derived stellar age ($>5$~Gyr), 
we do not include any activity contributions in the remaining of our analysis, except for an uncorrelated jitter term ($\sigma_\mathrm{jitter}$).

\section{Ruling out false positive scenarios}\label{sec:false_positives}
Previous experience with {\it Kepler} shows that 
candidates in multiple systems have a much lower probability of being false positives \citep{latham11, Lissauer2012}. Nevertheless, it is always appropriate to perform a series of
checks
in order to exclude the possibility of a false positive. \par
We notice that the star has a good astrometric {\it Gaia} DR2 solution \citep{GaiaColl2018}, with zero excess noise and a re-normalised unit weight error (RUWE) of $1.1$, indicating that the single-star model provides a good fit to the astrometric observations. 
This likely excludes the presence of a massive companion that could contribute to the star's orbital motion in the {\it Gaia} DR2 astrometry, a fact that agrees with the absence of long-term trends in our RVs (see Section~\ref{sec:RV_signals}). \par
Moreover,
the overall RV variation below $25$~\ms and the 
shape of the CCFs of our HARPS-N spectra exclude the eclipsing binary scenario, which would be the most likely alternative explanation for the USP planet. \par
A further confirmation comes from the speckle imaging on the Southern Astrophysical Research (SOAR) telescope that \cite{ziegler2020} performed on some of the {\it TESS} planet candidate hosts. According to their analysis (see Tables~$3$ and $6$ therein), no companion is detected around \starname\ (being the resolution limit for the star $0.041$~arcsec, and the maximum detectable $\Delta$mag at separation of $1$~arcsec $4.76$~mag).
Still, the $21$~arcsec {\it TESS} pixels and the few-pixels wide point spread function (PSF) 
can cause the light from 
neighbours over an arc-minute away to contaminate the target light curve.
In the case of neighbouring eclipsing binaries (EBs), 
eclipses can
be diluted and mimic shallow planetary transits. 
For example, events at $\sim1$~mmag level as in TOI-561.01 and TOI-561.03
can be mimicked by a nearby eclipsing binary within the {\it TESS} aperture
with a $0.5$\% eclipse, but no more than
$7$ magnitudes fainter.
This condition is not satisfied
in our case, 
as
the only three sources within $100$~arcsec from \starname\ 
are
all fainter than $T=19.25$~mag and at 
a
distance greater than $59$~arcsec, according to the {\it Gaia} DR2 catalogue.\par
\begin{figure}
\centering
  \includegraphics[bb=3 160 583 707, width=\columnwidth]{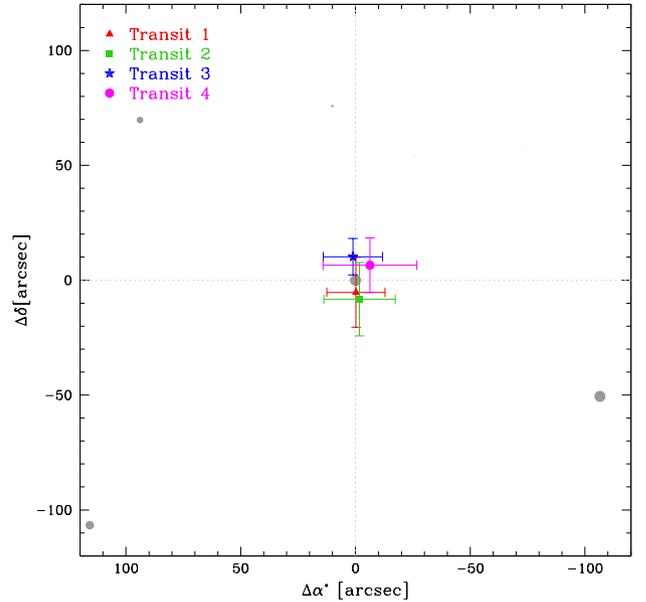}
   \caption{In-/out-of-transit difference centroid analysis of the transit events associated with the candidates TOI-561.01 (transit $2$ and $3$) and TOI-561.03 (transit $1$ and $4$).
   The star is centred at (0,0), and the grey circles are all the other stars in the {\it Gaia} DR2 catalogue, with dimension proportional to their apparent magnitude.}
      \label{fig:centroid}
\end{figure}

An independent confirmation was provided by the analysis of the in-/out-of-transit difference centroids on the {\it TESS} FFIs (Figure~\ref{fig:centroid}), adopting the procedure described 
in \cite{nardiello2020}.
The analysis of the in-/out-of transit stacked difference images confirms that, within a box of $10\times10$ pixels$^2$ ($\sim 200 \times 200$ arcsec$^2$) centred on TOI-561, the transit events associated with candidates .01 and .03 occur on our target star, while candidate .02 has too few in-transit points in the $30$-minute cadence images for this kind of analysis~---~
in any case,
its planetary nature 
will be
confirmed by the RV signal of \starname\ in Section~\ref{sec:system_architecture}. \par
Finally, in order to exclude the possibility that the transit-like features were caused by instrumental artefacts, we performed some additional checks on the light curve. 
We visually inspected the FFIs to spot possible causes
(including instrumental effects) 
inducing transit-like features, and we could not find any.
We re-extracted the short cadence light curve using the python package \texttt{lightkurve}\footnote{\url{https://github.com/KeplerGO/lightkurve}} \citep{2018ascl.soft12013L} with different photometric masks and apertures, and we corrected them by using the {\it TESS} Cotrending Basis Vectors (CBVs); 
the final results were in agreement with the {\it TESS}-released PDCSAP light curve.
We checked for systematics in every light curve pixel, and we found none.
Ultimately, we checked for correlations between the flux, the local background, the (X,Y)-position from the PSF-fitting, and the FWHM, with no results. 
Therefore, we conclude that all the transit-like features in the light curve are
real and
likely due to planetary transits.

\section{Data analysis tools}\label{sec:data_analysis_tools}

We performed the analysis presented in the next sections using \pyorbit\footnote{\url{https://github.com/LucaMalavolta/PyORBIT}, version 8.1} \citep{Malavolta2016, Malavolta2018}, a convenient wrapper for the analysis of transit light curves and radial velocities.

In the analysis of the light curve, for each planet we fitted the central time of transit ($T_0$), period ($P$),
planetary to stellar radius ratio ($R_\mathrm{p}/$\rstar ), and impact parameter $b$. 
In order to reduce computational time, we set a narrow, but still uninformative, uniform prior for period and time of transit, as defined by a visual inspection.
We fitted a common value for the stellar density \rhostar, imposing a Gaussian prior based on the value from Table~\ref{table:star_derived_params}. 
We included a quadratic limb-darkening law with Gaussian priors on the coefficients $u_1$, $u_2$,
obtained through a bilinear interpolation of limb darkening profiles by \cite{claret2018} \footnote{\url{https://vizier.u-strasbg.fr/viz-bin/VizieR?-source=J/A+A/618/A20}}.
We initially calculated the standard errors on $u_1$, $u_2$ using a Monte Carlo approach that takes
into account the errors on \teff\ and \logg\ as reported in Table~\ref{table:star_derived_params},
obtaining $u_1 = 0.393 \pm 0.007$ and $u_2 = 0.204 \pm 0.001$. We however decided to conservatively increase the error on both coefficients to $0.05$. In the fit we employed the parametrization ($q_1,\ q_2$) introduced by \cite{Kipping2013}.
Finally, we included a jitter term to take into account possible {\it TESS} systematics and short-term stellar activity noise. We assumed uniform, uninformative priors for all the other parameters, although the prior on the stellar density will inevitably affect the other orbital parameters.
All the transit models were computed with the {\tt batman} package \citep{Kreidberg2015}, with an exposure time of $120$ seconds and an oversampling factor of $10$ \citep{Kipping2010}.

In the analysis of the radial velocities, we allowed the periods to span between $2$ and $200$ days (\ie , the time span of our dataset) for the non-transiting planets,
while we allowed the semi-amplitude $K$ to vary between $0.01$ and $100$~\ms\ for all the candidate planets. These two parameters were explored in the logarithmic space.
For the transiting candidates, we used the results from the photometric fit (see Appendix~\ref{sec:appendix_transits})
to impose Gaussian priors on period and time of transit on RV analysis alone, while using the same uninformative priors as for the photometric fit when including the photometric data as well.

For all the signals except the USP candidate, we assumed eccentric orbits with a half-Gaussian zero-mean prior on the eccentricity (with variance $0.098$) according to \cite{vanEylen2019}, unless stated otherwise. 

We computed the Bayesian evidence using the {\tt MultiNest} nested-sampling algorithm \citep{Feroz2008,Feroz2009,Feroz2019} with the Python wrapper {\tt pyMultiNest} \citep{buchner2014}. 
In the specific case of the joint light curve and RV analysis (Section~\ref{sec:model_comparison}), we employed the \texttt{dynesty} nested-sampling algorithm \citep{skilling2004, skilling2006, speagle2020}, which allowed for the computation of the Bayesian evidence in a reasonable amount of time thanks to its easier implementation of the multi-processing mode.
We performed a series of test on a reduced dataset, and we verified that the two algorithms provided consistent results with respect to each other.
For all the analyses, we assumed $1000$ live points and a sampling efficiency of $0.3$, including a jitter term for each dataset considered in the model. 

Global  optimisation  of  the  parameters  was  performed using the differential evolution code 
{\tt PyDE}\footnote{\url{https://github.com/hpparvi/PyDE}}.  
The output parameters were used as a starting point for the Bayesian analysis performed with the {\tt emcee} package \citep{ForemanMackey2013}, a Markov chain Monte Carlo (MCMC) algorithm with an affine invariant ensemble sampler \citep{GoodmanWeare2010}.
We ran the chains with $2 n_{\mathrm{dim}}$ walkers, where $n_{\mathrm{dim}}$ is the dimensionality of the model, 
for a number of steps adapted to each fit,
checking the convergence with the Gelman-Rubin statistics \citep{GelmanRubin1992}, with a threshold value of $\hat{R} = 1.01$.  
We also performed an auto-correlation analysis of the chains: if the chains were longer  than $100$ times the estimated auto-correlation time and this estimate changed by less that $1\%$, we considered  the chains as converged. 
In each fit, we conservatively set the burn-in value as a number larger than the convergence point as just defined, and we applied a thinning factor of $100$.

\section{Unveiling the system architecture}\label{sec:system_architecture}
\subsection{Planetary signals in the RV data}\label{sec:RV_signals}
Before proceeding with a global analysis, we checked whether we could independently recover the signals identified by the {\it TESS} pipeline (Section~\ref{sec:tess_photometry}) in our RV data only.
The periodogram analysis 
of the RVs 
in Section \ref{sec:stellar_activity} highlighted the presence of several peaks
not related to the stellar activity. 
In particular, an iterative frequency search, 
performed subtracting at each step the frequency values previously
identified, supplied the frequencies $f_1 = 0.039$~d$^{-1}$ ($P_1 \simeq 25.6$~d), $f_2 = 0.006$~d$^{-1}$ or $0.013$~d$^{-1}$ ($P_2 \simeq 170$~d or $\simeq 78$~d) with the two frequencies being related to each other (\ie, removing one of them implies the vanishing of the other one), $f_3 = 0.093$~d$^{-1}$ ($P_3 \simeq 10.8$~d, corresponding to the TOI-561.01 candidate), and $f_4 = 2.239$~d$^{-1}$ ($P_4 \simeq 0.45$~d, corresponding to the TOI-561.02 candidate). After removing these four signals, no other clear dominant frequency emerged in the residuals.
Since any attempt to perform a fit of the RVs to characterise the transiting candidates without accounting for 
additional dominant
signals would lead to unreliable results, we decided to test the presence of additional planets in a Bayesian framework. 
We considered four models, the first one (Model 0)
assuming 
the three transiting candidates only,\ie, TOI-561.01, .02, .03, and then including an additional planet in each of the successive models,\ie, TOI-561.01, .02, .03 plus one (Model 1), two (Model 2) and three (Model 3) additional signals, respectively.
We computed the Bayesian evidence 
for each model 
using the {\tt MultiNest} nested-sampling algorithm,
following the prescriptions as specified in Section~\ref{sec:data_analysis_tools}.
We report the obtained values in Table~\ref{table:logZ}.
According to this analysis, we concluded that the model with two additional signals, \ie, Model 2 (with no trend), is strongly favoured
over the others, with a difference in the logarithmic Bayes factor $ 2 \, \Delta\ln\mathcal{Z} > 10$ \citep{kass&raftey1995}, 
both compared to the case with one or no additional signals. 
In the case of a third additional signal (Model 3),
the difference with respect to the two-signal model was less than $2$,
indicating that there was no strong evidence to favour this more complex model over the simpler model with two additional signals only \citep{kass&raftey1995}.
We repeated the analysis 
first including
a linear and then a quadratic trend in each of the four models.
In all cases, the Bayesian evidence systematically disfavoured the presence of any trend\footnote{For the model with three additional signals and a quadratic trend, the calculation of the Bayesian evidence did not converge.}.\par
The first additional signal was associated with a candidate with $f \simeq 0.04$~d$^{-1}$ ($P \simeq 25.6$~d), which corresponds to the strongest peak in the RVs periodogram. Concerning the second additional signal, the {\tt MultiNest} run highlighted the presence of two clusters of solutions, peaked at about $f = 0.013$~d$^{-1}$ or $0.013$~d$^{-1}$, \ie, $P = 78$ and $180$ days respectively.
The frequency analysis confirmed that the signals are aliases of each other, since when we subtract one of them, the other one also disappears. The alias peak is visible in the low-frequency regime of the spectral window (Figure~\ref{fig:activity_periodograms}, bottom panel). We should also consider that the longer period is close to the time baseline of our data.
In order to disentangle the real frequency from its alias, we computed the Bayesian evidence of the two possible solutions, first allowing the period to vary between $50$ and $100$ days, and then between $100$ and $200$ days.
The Bayesian evidence slightly favoured the solution with $P \sim 78$~d, even if not with strong significance ($\Delta\ln\mathcal{Z} \simeq 2$). 
Since we could not definitely favour one solution over the other, we decided to perform all the subsequent analyses using both sets of parameters.\par
\begin{figure}
\centering
\includegraphics[width=\linewidth]{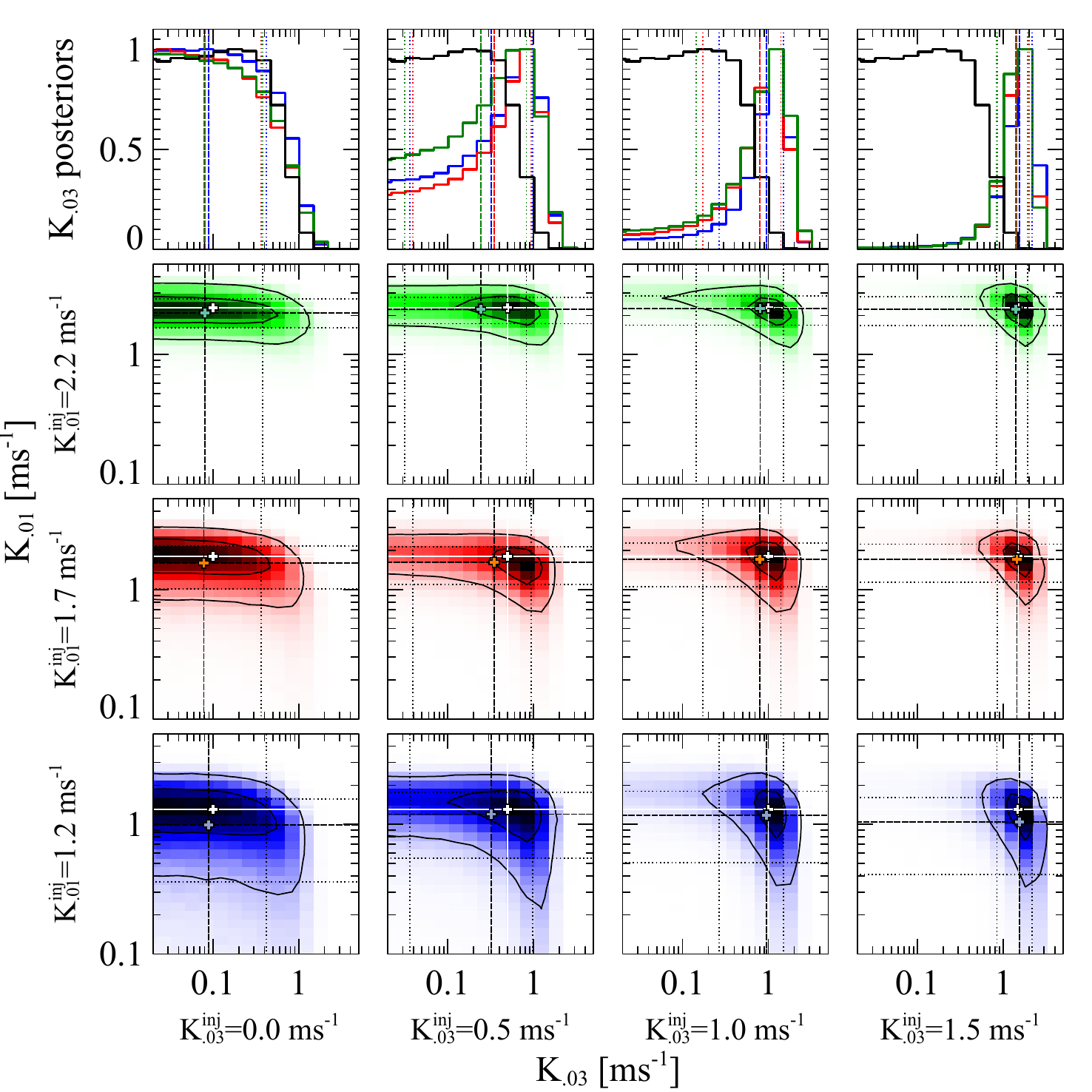}
  \caption{Posterior distributions (in the top panels, the blue, red and green lines respectively) of the retrieved RV signal of TOI-561.03 according to different injected values for the RV semi-amplitudes of candidates .01 and .03. The black line in the top panels corresponds
  to the observed posterior of the RV semi-amplitude of candidate .03. Median and 1-$\sigma$ values are marked with vertical dashed and dotted lines respectively. 
          }
      \label{fig:Kd_Kc_simulation}
\end{figure}

Another important outcome of our frequency search is the absence of a signal with a periodicity of $\sim 16$ days, that is, the transiting candidate TOI-561.03. 
Therefore, in order to test our ability to recover the planetary signals, we performed a series of injection/retrieval simulations, thoroughly explained in Appendix~\ref{sec:appendix_RV_model}. 
The results of this injection/retrieval test are summarised in Figure~\ref{fig:Kd_Kc_simulation}. We found that the injected RV amplitude of .01 is not significantly affecting the retrieved value for .03, \ie\ the cross-talk between the two signals is negligible. We verified that the same conclusion applies to the other signals as well. 
More importantly, any attempt to retrieve a null signal at the periodicity of 
the candidate planet
.03 would result in an upper limit of $\approx 0.5$~\ms\, as we actually observe with the real dataset, when exploring the $K$ parameter in logarithmic space. Any signal equal or higher than $1$~\ms\ would have been detected ($> 2 \sigma$), even if marginally. A signal with amplitude of $0.5$~\ms\ would not lead to the detection of the planet (intended as a 3-$\sigma$ detection), but the retrieved posterior is expected to differ substantially from the observed one, especially on the lower tail of the distribution. 
We conclude that the planetary candidate TOI-561.03 is undetected in our RV dataset, with an upper limit on the semi-amplitude of $0.5$~\ms (\mplanet~$< 2.0$~\mearth).

\begin{table}
  \caption{Logarithmic Bayesian evidences for the different models under exam. Model $0$ corresponds to the model with no additional RVs signal other than 
  the signals from
  the three transiting candidates \ie, TOI-561.01, .02, .03. Model $1$, $2$ and $3$ correspond to the models with the three transiting candidates plus one, two and three additional planets, 
  respectively. All the values are expressed with respect to Model $0$. We note that the reported errors, as obtained from the nested sampling algorithm, are likely underestimated \citep{Nelson2020}.}
\label{table:logZ}      % is used to refer this table in the text
\centering                                      % used for centering table
\begin{tabular}{c c c c c}          % centered columns 
\hline\hline                        % inserts double horizontal lines
& Model $0$ & Model $1$ & Model $2$ & Model $3$\\
\hline                                 % inserts single horizontal
$\ln\mathcal{Z}$ & $0.0 \pm 0.1$ & $13.4 \pm 0.2$ & $26.1 \pm 0.2$ & $28 \pm 0.2$\\\hline
\end{tabular}
\end{table}

\subsection{Transit attribution}\label{sec:transit_attribution}
\begin{figure}
\centering
\includegraphics[width=0.8\linewidth]{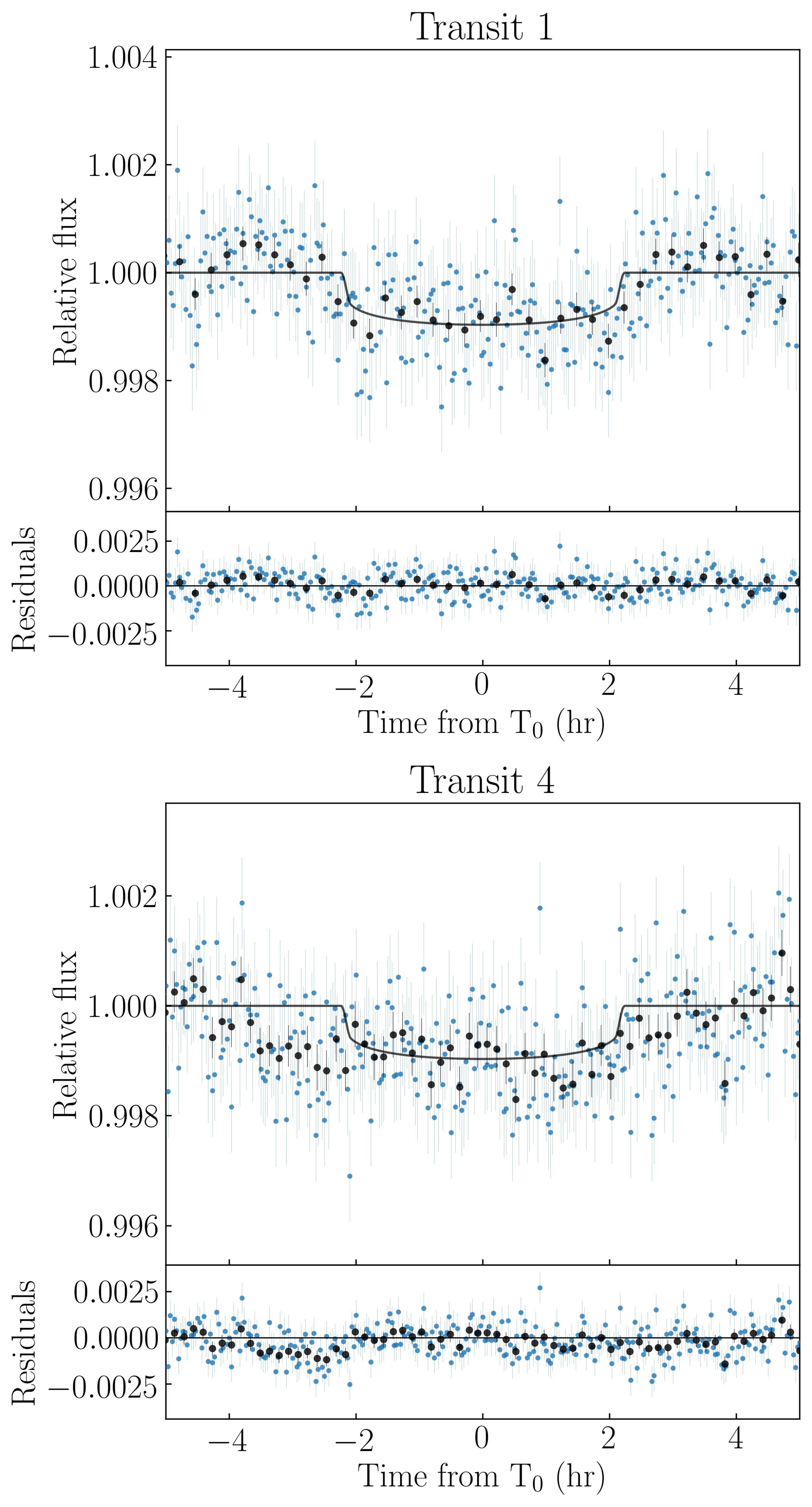}
  \caption{Transit $1$ ($ T_{0} \simeq 1521.9$~d) and $4$ ($T_{0} \simeq 1538.2$~d) in the {\it TESS} detrended light curve associated with the candidate TOI-561.03. The best-fitting transit model from the three-planet model photometric fit is over-plotted (black solid line). The  black
  dots are the data points binned over $15$ minutes. With respect to transit $1$, the duration of transit $4$ looks underestimated by the global model,
  with a systematic offset in the residuals, especially in the pre-transit phase.}
      \label{fig:transit_d_comparison}
\end{figure}

Given the non-detection of the planetary candidate TOI-561.03 in the RV data, 
we investigated more closely the transit-like features associated with this candidate in the {\it TESS} light curve, at $T_{0}$\footnote{All the $T_0$s in this section are expressed in BJD-2457000.} $\simeq 1521.9$~d and $T_{0} \simeq 1538.2$~d, referred from now on as transit $1$ and $4$
respectively, given their sequence in the TESS light curve (when excluding the transits of the USP candidate).
From our preliminary three-planet photometric fit (Figure~\ref{fig:photometry1}), we noted that, with respect to the other candidates, TOI-561.03 appears to have a longer transit duration compared to the model, and the residuals show some deviations in the ingress/egress phases.
To better understand the cause of these deviations, we checked how the model fits each transit. As Figure~\ref{fig:transit_d_comparison} shows, the global model appears to better reproduce the first transit associated with TOI-561.03 (transit $1$) than the second 
transit (transit $4$), that has a duration that looks underestimated by the model.
Moreover, a two-sample Kolmogorov--Smirnov statistical test\footnote{We used the Python version implemented in \texttt{scipy.stats.ks\_2samp}.}
\citep{Hodges1958} on the residuals of transit $1$ and $4$ suggests that the two residual samples are not drawn from the same distribution
(threshold level $\alpha=0.05$, statistics $\mathrm{KS}=0.178$, $p\mathrm{-value} \ll 0.01$).

\begin{figure}
\centering
\includegraphics[width=\linewidth]{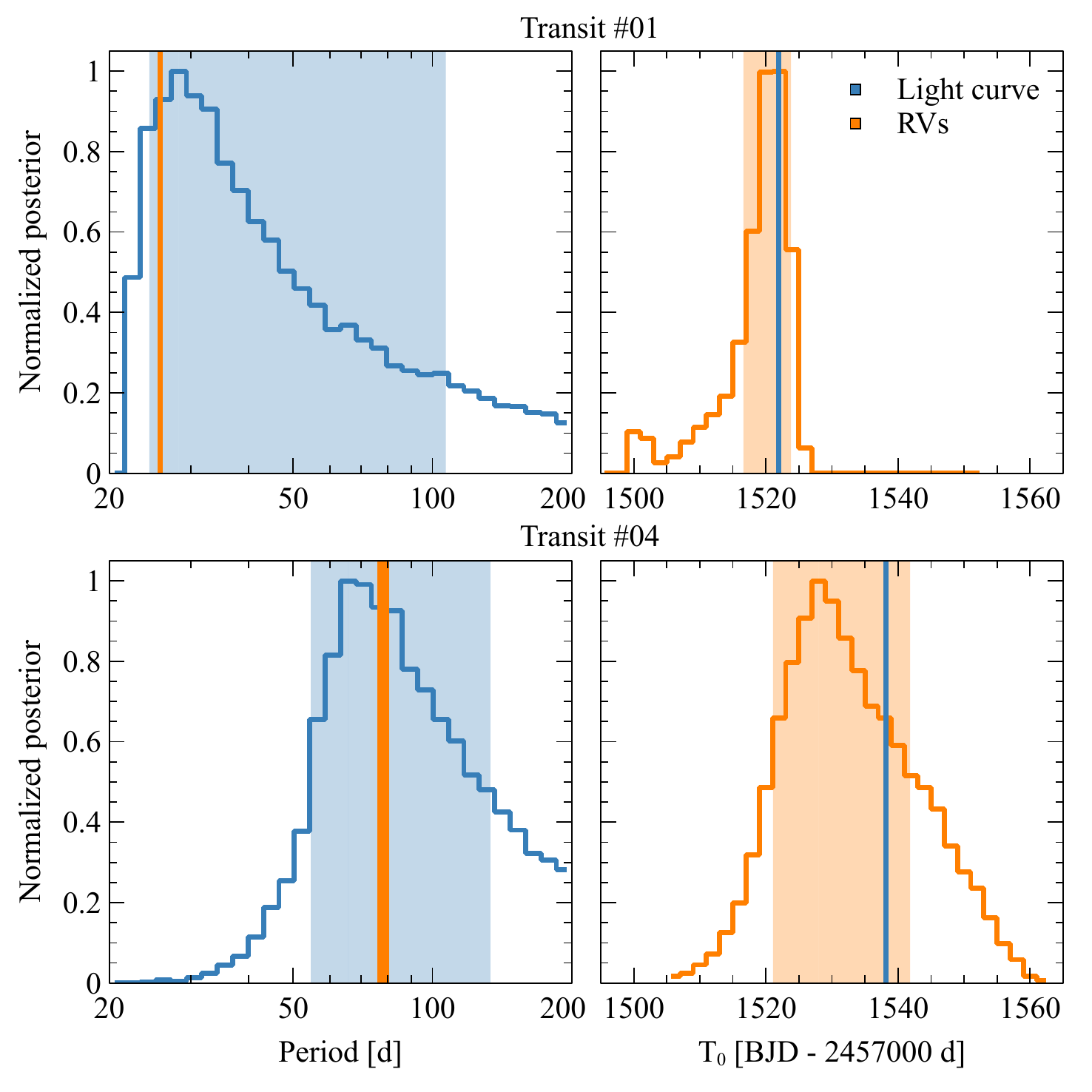}
  \caption{Comparison between period (left panels) and $T_0$ (right panels) obtained from the RV fit and from the fit of each single transit. Top and bottom panels refer to transit $1$ and $4$, respectively. Each panel shows the posterior distribution of the analysed parameter, and the shaded area indicates the region within the $68.27$-th percentile from the mode of the distribution. The vertical solid lines indicate the inferred best-fitting value of the parameter, with thickness proportional to the associated error.     
          }
      \label{fig:LC_RV_transit_attribution}
\end{figure}

Therefore, we hypothesised that the two transit-like features 
may be
unrelated, \ie,
they correspond to the transits of two distinct
planets. 
Since two 
additional
planets are actually detected in the RV data, and their periods are
longer than the {\it TESS} light curve interval (i.e., that {\it TESS} can detect, at most, only one transit for
each of them),
we tested the possibility that
the two transits previously associated with TOI-561.03 could indeed be due to the two additional planets inferred from the RV analysis.
To check our hypothesis, we first analysed the RV dataset with a model encompassing four planets, of which only .01 and .02 have period and time of transit constrained by {\it TESS}. 
In other words, we performed the same RV analysis as described in Appendix~\ref{sec:appendix_RV_model}, but without including TOI-561.03 in the model.
We repeated the analysis twice in order to disentangle the periodicity at $78$~d from its alias at $180$~d, and \textit{vice versa}. We used the posteriors of the fit to compute the expected time of transit of the outer planets. 
We then performed two independent fits of transit $1$ and $4$ with \pyorbit, following the prescriptions as specified in 
Section~\ref{sec:data_analysis_tools}.
We imposed a lower boundary on the period of $22$ days, in order to exclude the periods that would imply a second transit of the same planet in the {\it TESS} light curve, and an upper limit of $200$ days. 
As a counter-measure against the degeneracy between eccentricity and impact parameter in a single-transit fit, we kept the \cite{vanEylen2019} eccentricity prior knowing that high eccentricities for such a compact, old system are quite unlikely \citep{vanEylen2019}.
Finally we compared the posteriors of period and time of transit from the photometric fit with those from radial velocities, knowing that the former will 
provide
extremely precise transit times, but a broad distribution in period,
while RVs give us precise periods, but little information on the transit times.
The results are summarised in 
Figure~\ref{fig:LC_RV_transit_attribution}:
the $25.7 \pm 0.3$~d signal detected in the RVs is located in the vicinity of the main peak of transit $1$ period distribution, while the $78.6_{-2.5}^{+1.8}$~d signal is close to the main peak in transit $4$ period distribution.
Moreover, Figure~\ref{fig:LC_RV_transit_attribution} definitely confirms that 
both the conjunction times 
inferred from the RV fit corresponding to the $\sim 25$ and $\sim 78$ days signals, respectively $T_0 = 1520_{-6}^{+3}$~d and $T_0 = 1532_{-9}^{+12}$~d, are consistent with the (much more precise) $T_0$s inferred from the individual fit of transit $1$ ($T_0 = 1521.885 \pm 0.004$~d) and $4$ ($T_0 = 1538.178 \pm 0.006$~d) respectively. Regarding the alias at $182 \pm 7$ days, while the RV period is consistent with the corresponding posterior from the transit fit, 
the conjunction time 
$T_0 = 1628 \pm 13$~d that is derived from our analysis is not
compatible with any of the transits in the {\it TESS} light curve. 
We also note that the proportion of the orbital period covered by the {\it TESS} photometry is $\sim 2.3$ times larger for the candidate with $78$~d period, thus increasing the chance of getting a transit of it.
In conclusion, 
taking into account both photometric and RV observations, 
the most plausible solution
for the TOI-561 system is a
four-planet configuration in which 
transits
$1$ and $4$ are associated with the planets that have periods of $\sim 25$~d and $\sim 78$~d detected in the RV data, and the $180$~d signal is considered an alias of the $78$~d signal. \\
Given this
final
configuration,
hereafter
we will refer to the planets with period $\sim0.45$, $\sim10.8$, $\sim25$ and $\sim78$ days as planets b, c, d and e, respectively.

\subsection{The system architecture}\label{sec:final_solution}
\begin{figure*}
\centering
\includegraphics[width=\linewidth]{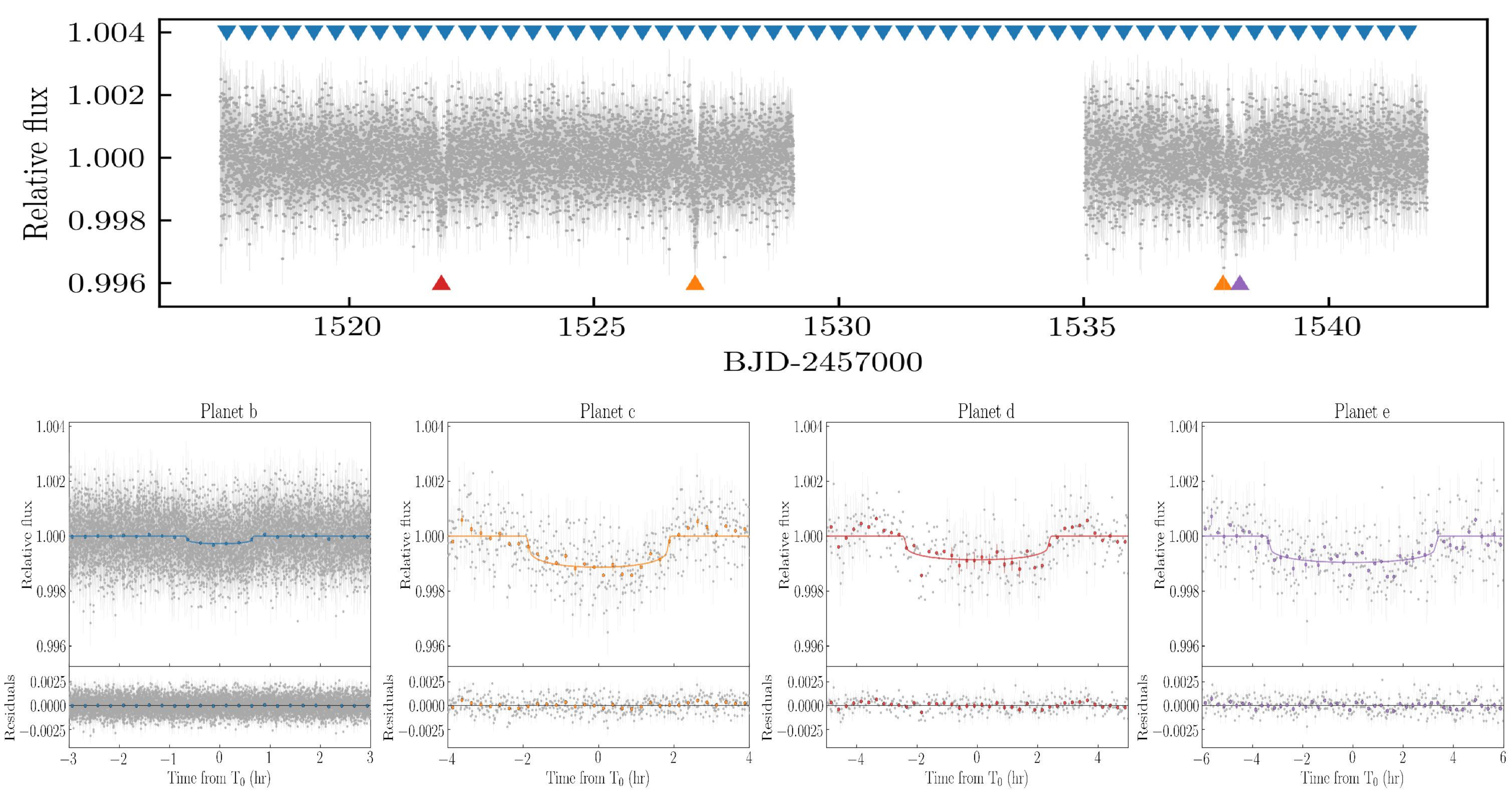}
  \caption{\emph{Top}: $2$-minute cadence flattened light curve of \starname. 
  The transits of planet b ($P \sim 0.45$~d), c ($P \sim 10.8$~d), d ($P \sim 25.6$~d), e ($P \sim 77.2$~d) are highlighted with blue, orange, red and purple triangles, respectively. \emph{Bottom}: \starname\ phase-folded $2$-minute light curves over the best-fitting models (solid lines) for the four planets.  The light curve residuals are shown in the bottom panel.
          }
      \label{fig:photometry2}
\end{figure*}

\begin{figure}
\centering
\includegraphics[height=0.8\textheight]{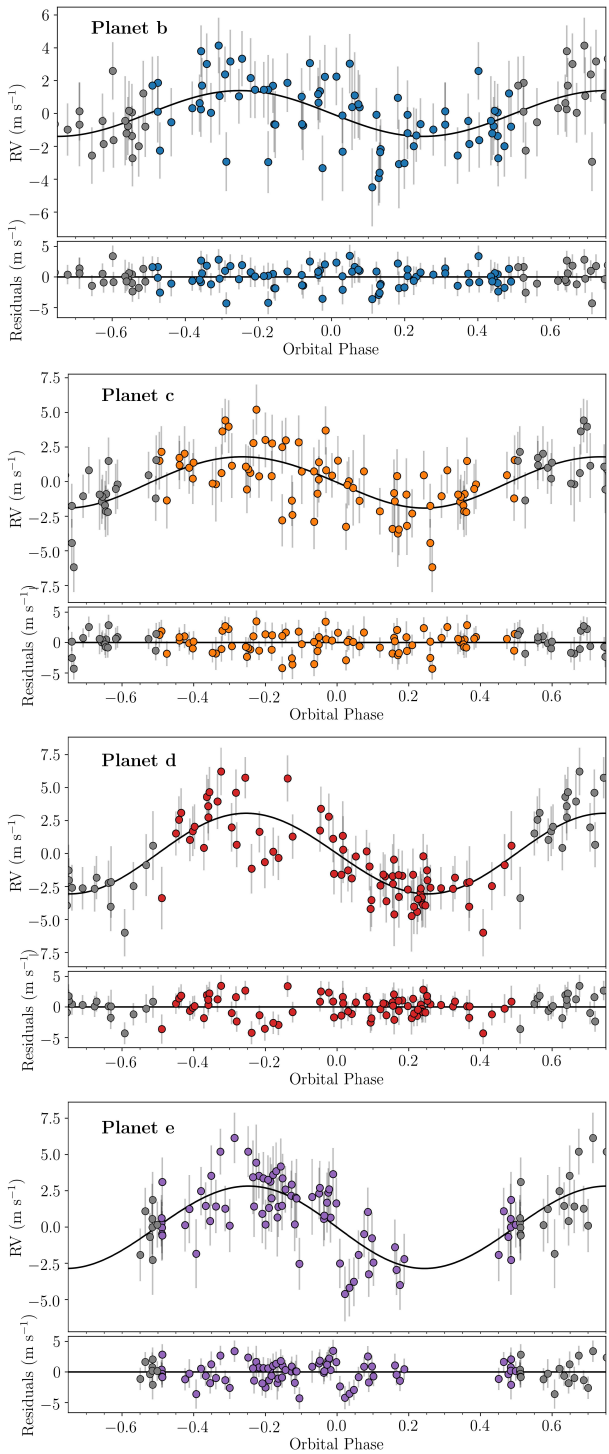}
  \caption{Phase-folded RV fit with residuals from the joint four-planet photometric and RV analysis. Planets b, c, d, and e are shown in blue, orange, red and purple, respectively. The reported errorbars include the jitter term, added in quadrature.}
      \label{fig:RV_phase}
\end{figure}

\begin{figure}
\centering
\includegraphics[width=\linewidth]{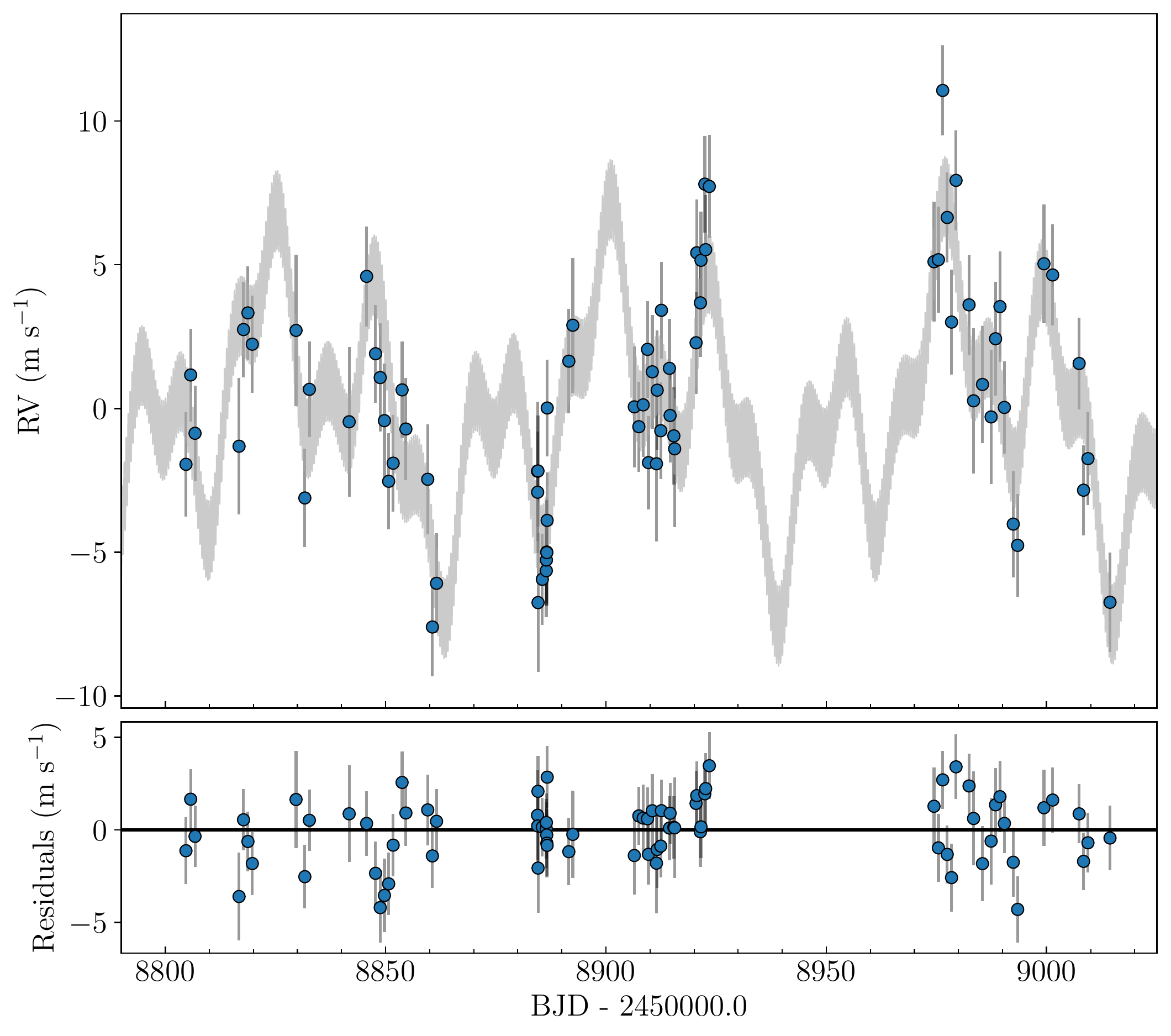}
  \caption{Four-planet model from the joint photometric and RV analysis. The grey curve is the the best-fitting model, and the blue points are the HARPS-N data. The residuals are shown in the bottom panel. The reported errorbars include the jitter term, added in quadrature.}
      \label{fig:RV_model}
\end{figure}

Given the presence of two single-transit planets in our data, a joint photometric and RV
modelling
is necessary in order to characterise the orbital parameters 
of all members of the TOI-561 system
in the best possible way.
We considered a four-planet model, with a circular orbit for the USP planet and allowing nonzero-eccentricity orbits for the others.
We performed the \pyorbit\ fit as specified in Section~\ref{sec:data_analysis_tools}, running the chains for $150\,000$ steps, and discarding the first $50\,000$ as burn-in.
We summarise
the results of our best-fitting model in Table~\ref{table:joint_parameters}, and 
show the transit models, the phase folded RVs, and the global RV model in Figures~\ref{fig:photometry2}, \ref{fig:RV_phase}, and \ref{fig:RV_model} respectively. 
We obtained a robust detection of the USP planet (planet b) RV semi-amplitude
($K_\mathrm{b} = 1.39 \pm 0.32$~\ms), that corresponds to a mass of $M_\mathrm{b} = 1.42 \pm 0.33$~\mearth,
while for the $10.8$~d period planet (planet c) we obtained $K_{\rm c} = 1.84 \pm 0.33$~\ms, corresponding to $M_{\rm c} = 5.40 \pm 0.98$~\mearth. We point out that the here reported value of $K_\mathrm{b}$ and $M_\mathrm{b}$ is obtained from the joint photometric and RV fit. However, the final value of $K_\mathrm{b}$ and $M_\mathrm{b}$ that we decided to adopt (see Section~\ref{sec:USP_nighlty_offset} for more details) is the weighed mean between the values obtained from the joint fit reported in this section and from the floating chunk offset method described in the next section.
In addition, we inferred the presence of two additional planets, with periods 
of $25.62 \pm 0.04$ days (planet d) and $77.23 \pm 0.39$ days (planet e),
and robustly determined semi-amplitudes of $K_{\rm d} = 3.06 \pm 0.33$~\ms ($M_{\rm d}= 11.95 \pm 1.28$~\mearth)
and $K_{\rm e} = 2.84 \pm 0.41$~\ms ($M_{\rm e} = 16.0 \pm 2.3$~\mearth). 
Both planets show a single transit in the {\it TESS} light curve, previously attributed to a transiting planet with period $\sim16$~d, whose presence has however been ruled out by our analysis. 
This allowed us to infer a planetary radius of $R_{\rm d} = 2.53 \pm 0.13$~\rearth\ and $R_{\rm e} = 2.67 \pm 0.11$~\rearth\ for planet d and e respectively. \par
We performed the stability analysis of our determined solution, computing the orbits for $100$~Kyr with 
the \texttt{whfast} integrator (with fixed time-step of $0.1$~d)
implemented within the \texttt{rebound} package
\citep{rein2012, Rein2015MNRAS.452..376R}.
During the integration we checked the dynamical stability of the solution with
the Mean Exponential Growth factor of Nearby Orbits (MEGNO or \megno) indicator
developed by \citet{CincottaSimo2000} and
implemented within \texttt{rebound} by \citet{Rein2016MNRAS.459.2275R}.
We ran $10$ simulations with initial parameters drawn from a Gaussian distribution
centred on the best-fitting parameters and standard deviation
derived in this section. All the $10$ runs
resulted in a MEGNO value of $2$,
indicating that the family of solutions is stable. \par
Finally, we checked the presence of any additional signal in
the RVs residuals after removing the four-planet model contribution. 
The GLS periodogram showed a non-significant peak at $\sim 2.5$ days, with a normalised power of $0.20$, that is, below the $1\%$ FAP threshold ($0.26$). 
As a supplemental confirmation, we ran a \pyorbit\ fit of the RVs, assuming first a four-planet model plus an additional signal, and then a four-planet model adding a Gaussian Process (GP) regression.
For the latter approach, we employed the quasi-periodic kernel as formulated by \cite{Grunblatt2015}, with no priors on the GP hyper-parameters, since we could not identify any activity-related signal in the ancillary datasets (see Section~\ref{sec:stellar_activity})\footnote{We are well aware that this is a sub-optimal use of GP regression, and that this approach may be justified in this specific case only as an attempt to identify additional signals.}. In both cases, the (hyper-)parameters of the additional signal did not reach convergence, while the results for the four transiting planets were consistent with those reported above. \\
Considering these results, we adopt the parameters and configuration determined in this section as the representative ones for the \starname\ system, with the only exception of the mass and semi-amplitude of TOI-561 b, that we discuss in the next section.

\begin{table*}
\caption{Final parameters of the TOI-561 system.}
\label{table:joint_parameters}     
\centering
\begin{tabular}{l c c c c} 
  \hline\hline                        
  Parameter & TOI-561b & TOI-561c & TOI-561d & TOI-561e\\
  \hline
  $P$ (d) & $0.446578 \pm 0.000017$& $10.779 \pm 0.004$& $25.62 \pm 0.04$ & $77.23 \pm 0.39$ \\
  $T_0^{\rm a}$ (d) & $1517.498 \pm 0.001$ & $1527.060 \pm 0.004$  & $1521.882 \pm 0.004$ & $1538.181 \pm 0.004$\\
  $a/$\rstar & $2.646 \pm 0.031$ & $22.10 \pm 0.26$& $39.35 \pm 0.46$ & $82.13 \pm 0.99$ \\
  $a$ (AU)  & $0.01055 \pm 0.00008$ & $0.08809 \pm 0.0007$ & $0.1569 \pm 0.0012$ & $ 0.3274_{-0.0027}^{+0.0028}$\\
  $R_\mathrm{p}/$\rstar & $0.0152 \pm 0.0007$ & $0.0308 \pm 0.0009$& $0.0271 \pm 0.0014$ &  $0.0286 \pm 0.0011$\\
  \rplanet\ (\rearth) & $1.423 \pm 0.066$& $2.878 \pm 0.096$ & $2.53 \pm 0.13$ & $2.67 \pm 0.11$ \\
  $b$ & $0.14_{-0.10}^{+0.13}$ & $0.18_{-0.12}^{+0.16}$ & $0.32_{-0.19}^{+0.17}$ & $0.34_{-0.20}^{+0.13}$\\
  $i$ (deg) & $87.0_{-2.8}^{+2.1}$ & $89.53_{-0.39}^{+0.32}$ & $89.54_{-0.21}^{+0.28}$ & $89.75_{-0.08}^{+0.14}$\\
  $T_{14}$ (hr) & $1.327_{-0.030}^{+0.021}$ & $3.77_{-0.15}^{+0.07}$& $4.85_{-0.35}^{+0.20}$ & $6.96_{-0.38}^{+0.34}$ \\
  $e$ & $0$ (fixed) & $0.060_{-0.042}^{+0.067}$& $0.051_{-0.036}^{+0.064}$ & $0.061_{-0.042}^{+0.051}$ \\
  $\omega$ (deg) & $90$ (fixed) & $200_{-49}^{+55}$ & $246_{-124}^{+67}$ & $155 \pm 83$\\
  $K^{\rm b}$ (\ms) &$1.56 \pm 0.35$ & $1.84 \pm 0.33$& $3.06 \pm 0.33$ & $2.84 \pm 0.41$ \\
  \mplanet$^{\rm b}$\ (\mearth) &$1.59 \pm 0.36$ & $5.40 \pm 0.98$& $11.95 \pm 1.28$ & $16.0 \pm 2.3$\\
  \rhoplanet\ (\rhoearth) & $0.55 \pm 0.14$& $0.23 \pm 0.05$& $ 0.74 \pm 0.14$& $0.84 \pm 0.16$ \\
  \rhoplanet\ (\gcm) & $3.0 \pm 0.8$ & $1.3 \pm 0.3$& $4.1 \pm 0.8$ & $4.6 \pm 0.9$\\
 \hline 
 \multicolumn{5}{c}{\emph{Common parameter}}\\
 \hline
  \rhostar\ (\rhosun) & $1.248 \pm 0.043$ & & & \\  
  $u_1$ & $0.401 \pm 0.048$ & \\
  $u_2$ & $0.208 \pm 0.049$ & \\
  $\sigma_{\rm jitter, ph}^{\rm c}$ & $0.000024_{-0.000011}^{+0.000018}$\\
  $\sigma_{\rm jitter}^{\rm d}$ (\ms) & $1.29 \pm 0.23$ & & & \\
  $\gamma^{\rm e}$ (\ms) & $79702.58 \pm 0.29$ & & & \\
 \hline
\multicolumn{5}{l}{$^a$ BJD$_{\rm TDB}$-2457000.}\\
\multicolumn{5}{l}{\makecell[l]{$^b$ The here reported values of planet b correspond to the weighted mean between the values inferred \\ from the floating chunk offset method ($K_{\rm b} = 1.80 \pm 0.38$~\ms, $M_{\rm b} = 1.83 \pm 0.39$~\mearth) and from\\ the joint photometric and RV fit ($K_{\rm b} =  1.39 \pm 0.32 $~\ms, $M_{\rm b} = 1.42 \pm 0.33$~\mearth).}}\\
\multicolumn{5}{l}{$^c$ Photometric jitter term. $^d$ Uncorrelated RV jitter term. $^e$ RV offset.}\\
\end{tabular}
\end{table*}

\subsection{Alternative characterisation of the USP planet}\label{sec:USP_nighlty_offset}
\begin{figure}
\centering
\includegraphics[width=\linewidth]{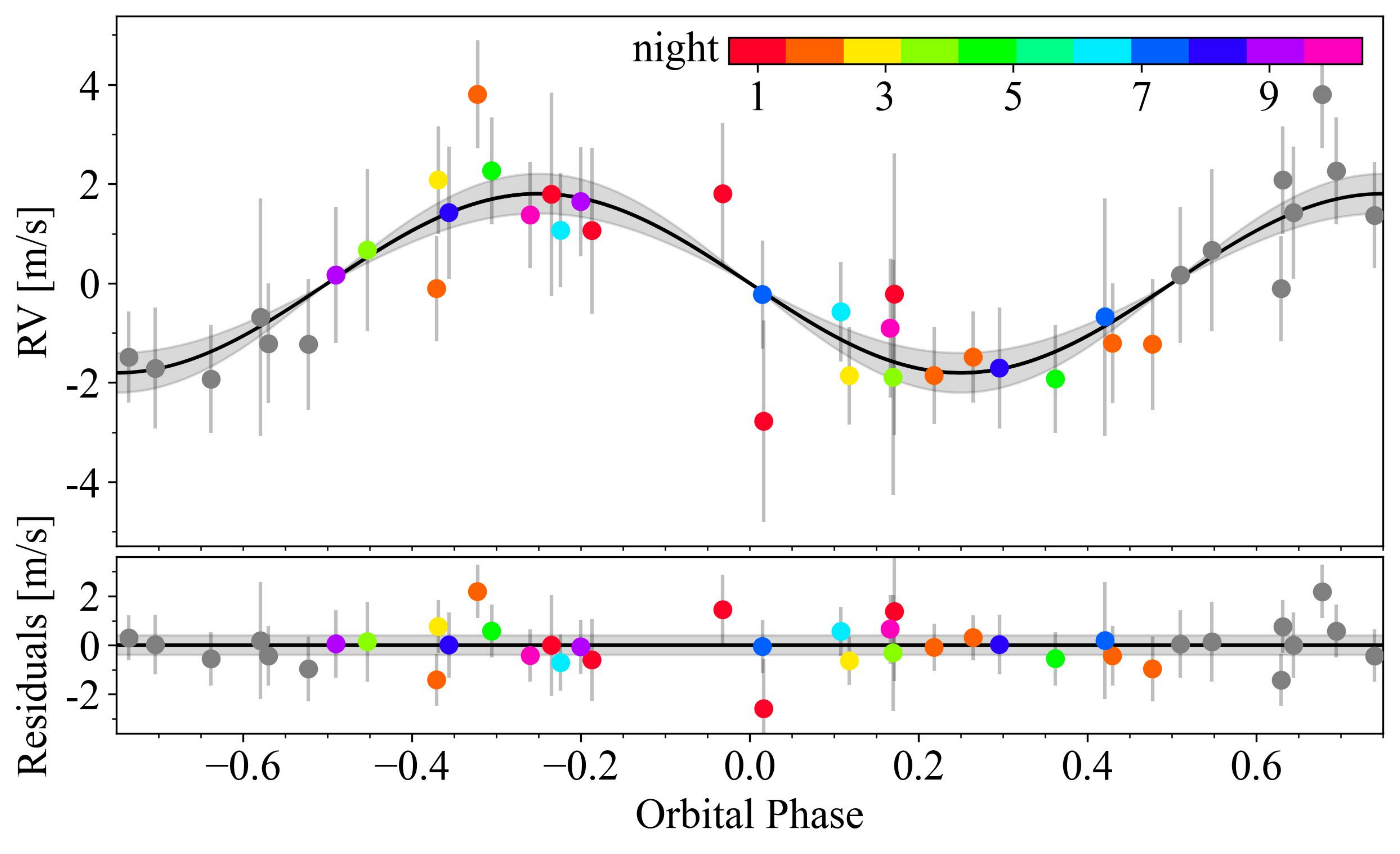}
  \caption{Phase folded RVs of the ten nights used to model the RV semi-amplitude of the USP 
  planet using the FCO approach.
          }
      \label{fig:USP}
\end{figure}

If the separation between the period of the planet and all the other periodic signals is large enough, and the RV signal has a similar or larger semi-amplitude, it is possible to determine the RV semi-amplitude for an USP planet without any assumptions about the number of planets in the system or the activity of the host star.
Under such conditions, during a single night, the influence of any other signal is much smaller than the measurement error and thus it can be neglected.
If two or more observations are gathered during the same night and they span a large fraction of the orbital phase, 
the RV semi-amplitude of the USP planet can be precisely measured by just applying nightly offsets to remove all the other signals (e.g. \citealt{hatzes2010, Howard2013, Pepe2013, Frustagli2020} for a recent example).
Such an approach, also known as floating chunk offset method (FCO; \citealt{Hatzes2014}), has proven extremely reliable even in the presence of complex activity
signals,
as shown by \cite{Malavolta2018}. 
In our case, the shortest, next periodic signal (\ie , TOI-561 c  at $10.78$ days) is $\simeq 24$ times the period of TOI-561 b (\ie , the USP planet at $0.45$ days), with similar predicted RV semi-amplitude, making this target suitable for the FCO approach. Thanks to our observational strategy (see Section \ref{sec:harpsn_RV}) 
we could use
ten different nights for this analysis. Most notably, during two nights we managed to gather six observations spanning nearly $5$ hours, \ie , more than $40$\% of the orbital period
of TOI-561 b,
at opposite orbital phases, thus providing a good coverage in phase of the RV curve. 
We did not include RV measurements with an associated error greater than $2.5$~\ms\ (see Appendix~\ref{sec:appendix_removal}).
We performed the analysis with \texttt{PyORBIT} as specified in Section~\ref{sec:data_analysis_tools}, assuming a circular orbit for the USP planet and including a RV jitter as a free parameter to take into account possible short-term stellar variability and any underestimation of the errorbars.
From our analysis, we obtained a RV semi-amplitude of $K_\mathrm{p} = 1.80 \pm 0.38$~\ms , corresponding to a mass of \mplanet~$= 1.83 \pm 0.39 $~\mearth . The resulting RV jitter is
$j < 0.9$~\ms ($84.13$-th percentile of the posterior).
We show the phase folded RVs of the USP planet in Figure~\ref{fig:USP}. \\
Since the greater reliability of this method over a full fit of the RV dataset is counter-balanced by the smaller number of RVs, we decided not privilege one over the other. Therefore, we assumed as final semi-amplitude and mass of TOI-561 b the weighted mean of the values obtained from the two methods (FCO approach and joint photometric and RV fit), \ie\ $K_\mathrm{b} = 1.56 \pm 0.35$~\ms, corresponding to a mass of $M_\mathrm{b} = 1.59 \pm 0.36$~\mearth. Table~\ref{table:joint_parameters} lists the above-mentioned values for TOI-561 b.

\section{Comparison with other models}\label{sec:model_comparison}
Our final configuration is quite different from the initial one suggested by the {\it TESS} automatic pipeline.
However, the analyses performed on the currently available data clearly disfavour the scenario with a $\sim 16$~d period candidate. 
In fact, in addition to the previous analyses, we also performed a joint photometric and RV fit assuming a five-planet model including the $16$~d period candidate, and assuming that the two additional signals seen in the RVs were caused by two non-transiting planets, the inner one with period of $\sim 25$~d and the outer one both in the case of $\sim 78$~d and $\sim 180$~d period. 
Such a model, including the TOI-561.01, .02, .03 candidates plus two additional signals, corresponds to the favoured model (Model $2$) identified in Section~\ref{sec:RV_signals}, and is therefore representative of the best-fitting solution when assuming the {\it TESS} candidate attribution.
In fact, Table~\ref{table:logZ} suggests that in this case two additional signals need to be added to the three transiting candidates to best reproduce the RV dataset, and therefore the five-planet model should be considered also in the joint photometric and RV modelling.

According to the Bayesian evidence (Table~\ref{table:logZ_joint}), computed with the \texttt{dynesty} algorithm as specified in Section~\ref{sec:data_analysis_tools}, the four-planet model is strongly favoured with respect to the five-planet model in both cases, with a difference in the logarithmic Bayes factor $ 2 \, \Delta\ln\mathcal{Z} \gg 10$ \citep{kass&raftey1995}. \par

Moreover, we checked the stability of the five-planet model solutions as described in Section~\ref{sec:final_solution}, with the external planet both on an orbit of $78$~d and $180$~d. 
For all the planetary parameters, including the mass of the $16$~d period planet\footnote{The mass of the $16$~d period planet obtained from the fit was $0.62 \pm 1.03$~\mearth\ and $1.19 \pm 1.27$~\mearth\ for the $\sim 78$~d and $\sim 180$~d external planet period, respectively. 
Obviously, when selecting the $10$ samples, the mass was constrained to positive values.}, we used the values and standard deviations derived from the joint photometric and RV fit, except for the inclination of the two external planets, that we fixed to $90$\degr. 

All of $10$ runs yielded unstable solutions, with a close encounter or an ejection occurring within the integration time.
In order to assess the origin of the instability of the system, we tested a four-planet configuration following the same procedure as above, removing one planet each time.
We found that the orbital configuration of the system could be stable
only if we remove the candidate with period of $\sim 16$~d.
Therefore, the stability analysis additionally confirms our determined four-planet configuration, ruling out the presence of a $\sim 16$~d period planet.

\begin{table}
  \caption{Logarithmic Bayesian evidences for the models considered in Section~\ref{sec:model_comparison}. Model $0$ corresponds to the four-planet model, that includes TOI-561.01, .02 and the two additional planets identified in the RVs, showing a single transit each. Model $1$ and $2$ correspond the five-planet model, \ie, including TOI-561.01, .02, .03 and the two additional RV planets (assumed in this case not to transit), in the case of an outer planet at $\sim 78$~d and $\sim 180$~d period respectively (see Section~\ref{sec:RV_signals}). All the values are expressed with respect to Model $0$. We note that the reported errors, as obtained from the nested sampling algorithm, are likely underestimated \citep{Nelson2020}}.
\label{table:logZ_joint}      % is used to refer this table in the text
\centering                                      % used for centering table
\begin{tabular}{c c c c}          % centered columns 
\hline\hline                        % inserts double horizontal lines
& Model $0$ & Model $1$ & Model $2$ \\
\hline                                 % inserts single horizontal
$\ln\mathcal{Z}$ & $0.0 \pm 0.9$ & $-77.8 \pm 1.0$ & $-76.9 \pm 1.0$ \\\hline
\end{tabular}
\end{table}

\section{Discussion and Conclusions}\label{sec:discussion_conclusions}
According to our analysis, \starname\ hosts four transiting planets, including an USP planet, a $\sim 10.8$~d period planet and two external planets with periods of $\sim 25.6$ and $\sim 77.2$ days.
The latter
were initially detected in the RVs data only, but based on our subsequent analyses we were able to identify a single transit of each planet in the {\it TESS} light curve; those transits were initially associated with a candidate planet with period of $\sim 16$~d, whose presence we ruled out. 
As a \lq lesson learned\rq, we would suggest that caution should be taken when candidate planets, detected by photometric pipelines, are based on just two transits. In such cases, one should not hesitate to consider alternative scenarios. \par 
\starname\ joins the sample of $88$ confirmed systems with $4$ or more planets\footnote{According to the \href{NASA exoplanet archive}{https://exoplanetarchive.ipac.caltech.edu/}.}, and it is one of the few multi-planet systems with both a mass and radius estimate for all the planets.
Our global photometric and RV model allowed us to determine the masses and densities of all the planets with high precision, with a significance of $\sim 4.4 \sigma$ for planet b and $> 5\sigma$ for planets c, d and e.
In Figure~\ref{fig:mass-radius} we show the position of \starname\ b, c, d and e in the mass-radius diagram of exoplanets with masses and radiii measured with a precision better than $30\%$. 
The comparison with the theoretical mass-radius curves
excludes an Earth-like composition ($\sim 33\%$ iron and $67\%$ silicates)
for 
all planets in the system, whose internal structure we further analyse in the following sections.

\subsection{TOI-561 b}\label{sec:planet_b}

The density ($\rho_{\rm b} = 3.0 \pm 0.8$~\gcm) of the USP planet is consistent with a $50\%$ (or even more) water composition. 
Such a composition may be compatible with a water-world scenario, where \lq water worlds\rq\ are 
planets
with massive water envelopes, in the form of high pressure H$_2$O ice, comprising $>5\%$ of the total mass. 
Even assuming the higher mass value inferred with the FCO method
($M_{\rm b} = 1.83 \pm 0.39$~\mearth, implying a density of $\rho_{\rm b} = 3.5 \pm 0.9$~\gcm),
TOI-561 b would be located close to the $25\%$ water composition theoretical curve in the mass-radius diagram, and it would be consistent with a rocky composition only at a confidence level greater than $ 2\sigma$ in both radius and mass.
Given its proximity to the host star (incident flux $F_\mathrm{p} \simeq 5100~F_{\oplus}$), 
the presence of any thick H-He envelope has to be excluded due the photo-evaporation processes that such old close-in planets are expected to suffer \cite[e.g.][]{Lopez2017}. 
Nevertheless, the possibility of a water-world scenario is an intriguing one. An H$_2$O-dominated composition would imply that the planet formed beyond the snow line, accreted a considerable amount of condensed water, and finally migrated inwards
\citep{Zeng2019}. 
While the determination of the precise interior composition of \starname\ b is beyond the scope of this work, if such an interpretation is proven trustworthy by future observational campaigns, \starname\ b would support the hypothesis that the formation of super-Earths with a significant amount of water is indeed possible.
However, an important caveat should be considered while investigating this scenario. 
If TOI-561 b was a water world, being more irradiated than the runaway greenhouse irradiation limit, the planet would present a massive and very extended steam atmosphere. Such an atmosphere would substantially increase the measured radius compared to a condensed water world \citep{turbet2020}. 
Therefore, a comparison with the condensed water-world theoretical curves should be used with caution, since in this case it could lead to an overestimation of the bulk water content \citep{turbet2020}. \par
\begin{figure}
\centering
\includegraphics[width=\linewidth]{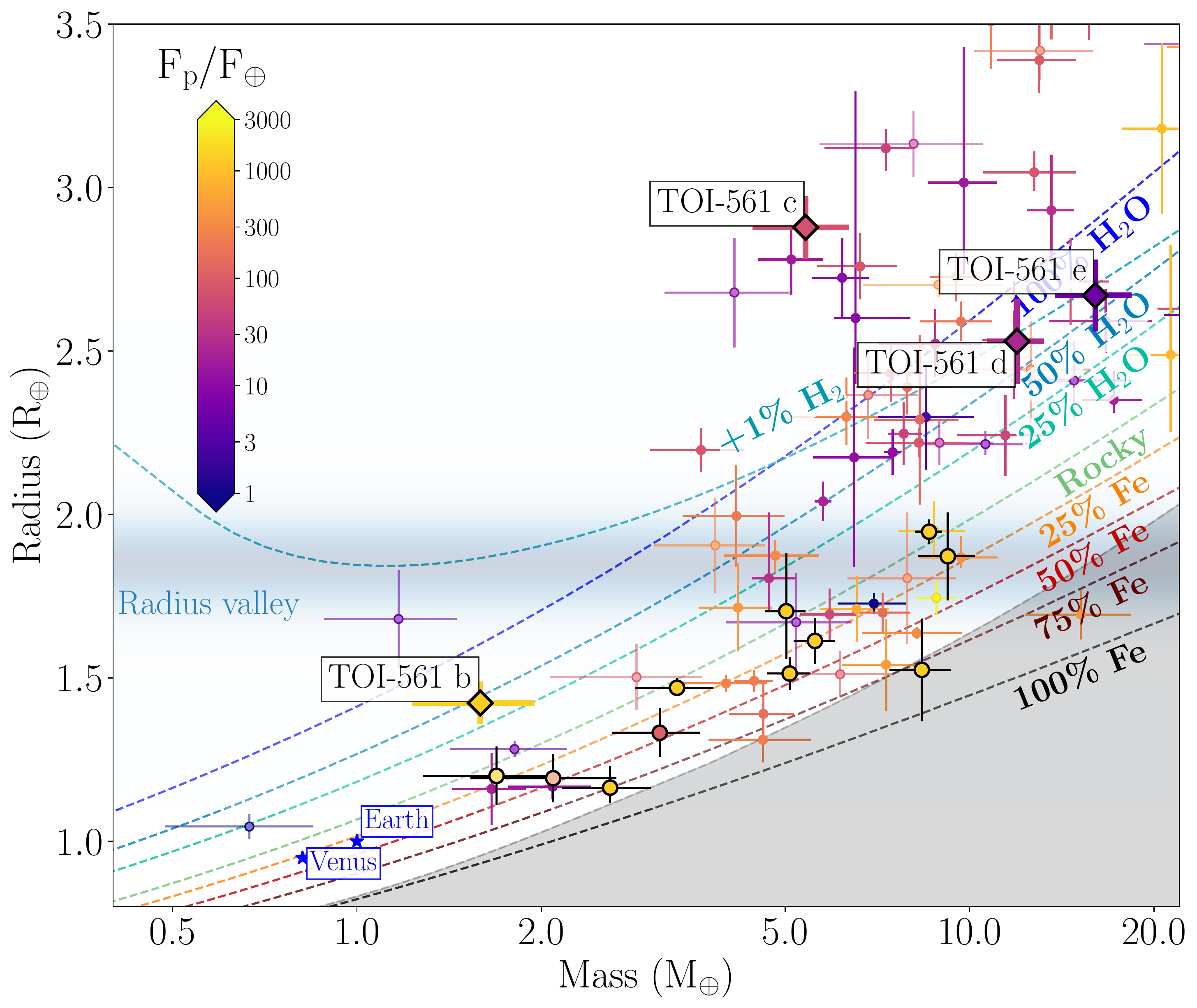}
\caption{Mass-radius diagram for known exoplanets with mass and radius measurements more precise than $30$\%, colour-coded according to their incidental flux in Earth units. The TOI-561 planets are labelled and represented with coloured diamonds. The USP planets are highlighted with black thick contours. The solid coloured lines represent the theoretical mass-radius curves for various chemical compositions according to \citet{Zeng2019}.
The shaded grey region marks the maximum value of iron content predicted by collisional stripping \citep{marcus2010}. 
The planetary data are taken from the The Extrasolar Planets Encyclopaedia catalogue (\url{http://exoplanet.eu/catalog/}) 
updated to August 17, 2020.}
\label{fig:mass-radius}
\end{figure}

Finally, we note that the USP planet is located on the opposite side of the radius valley, \ie\ the gap in the distribution of planetary radii at $\sim 1.7$-$2$~\rearth\ \citep{Fulton2017}, with respect to all the other planets in the system.  
The origin of the so-called radius valley is likely due to a transition between rocky and non-rocky planets with extended H-He envelopes, with several physical mechanisms proposed as explanation, i.e. photoevaporation \citep{Chen2016,owen2017, lopez&rice2018, Jin_2018}, core-powered mass loss \citep{ginzburg2018, gupta2019}, or superposition of rocky and non-rocky planet populations \citep{Lee2016, lopez&rice2018}. 
In the \starname\ system, planet c is located above the radius valley and it indeed appears to require a thick H-He envelope (see next section). In the same way, the compositions of planet d and e are consistent with the presence of a gaseous envelope. However, the density of \starname\ b is lower than expected for a planet located below the radius valley, where we mainly expect rocky compositions. 
Moreover, \starname\ b is the first USP planet with such a low measured density (see Figure~\ref{fig:mass-radius}). 
We note that also the USP planets WASP-$47$ e and $55$ Cnc e are less dense than an Earth-like rocky planet, even if both of them have higher densities than TOI-561 b, \ie,
$\rho_{\rm W47_e} = 6.4 \pm 0.6$~\gcm \citep{Vanderburg2017} and
$\rho_{\rm 55Cnc_e} = 6.3 \pm 0.8$~\gcm \citep{Demory2016} 
respectively. \cite{Vanderburg2017} proposed the presence of water envelopes as a possible explanation for the low densities of these two planets, even though the inferred amount of water was smaller than the one required to explain TOI-561 b location in the mass-radius diagram.
It should also be considered that both planets are more massive than TOI-561 b, \ie, $M_{\rm W47_e} = 6.83 \pm 0.66$~\mearth\ \citep{Vanderburg2017} and $M_{\rm 55Cn_e} = 8.08 \pm 0.31$~\mearth\ \citep{Demory2016}, thus increasing their chances of retaining a small envelope of high-metallicity volatile materials (or water steam) that could explain their low densities \citep{Vanderburg2017}. Given its smaller mass, this scenario is less probable for TOI-561 than for WASP-$47$ e and $55$ Cnc e, making the object even more peculiar. 
With its particular properties, this planet could be an intriguing case to test also other extreme planetary composition models.
For example, given the metal-poor alpha-enriched host star, the planet
is likely to have a lighter core composition.

\subsection{TOI-561 c, d and e}\label{sec:planet_c}

\starname\ c, with a density of $\rho_{\rm c} \sim 1.3$~\gcm, is located above the threshold of a $100\%$ water composition,
and given its position in the mass-radius diagram we suppose the presence of a significant gaseous envelope surrounding an Earth-like iron core and a silicate mantle, and possibly a significant water layer (high-pressure ice).
If the inner USP planet is water-rich, there is no simple planet formation scenario in which the outer three planets are
water-poor. It is simpler to assume that all four planets were
formed with similar volatile abundances, and that the inner USP
planet lost all of its H-He layer, plus much of its water content,
while the outer planets could keep them.
Following \citet{Lopez_2014}, assuming a rocky Earth-like core and a solar composition H-He envelope, we estimate that an H-He envelope comprising $\sim 4.9\%$ of the planet mass could explain the density of \starname\ c, using our derived stellar and planetary parameters. \par
Planets \starname\ d and e are consistent with a $>50$\% water composition, a feature that may place them among the water worlds. 
However, such densities are also consistent with the presence of a rocky core plus water mantel surrounded by a gaseous envelope. 
We estimate that a H-He envelope of $\sim 1.8$\% and $\sim 2.3$\% of the planet mass could explain the observed planetary properties. 

\subsection{Dynamical insights}\label{sec:dyn_conclusions}

Our analysis shows that the orbital inclinations of planets c, d and e are all consistent within $1\sigma$ (see Table~\ref{table:joint_parameters}), and 
that the difference with the inclination of the USP planet is of the order of $\Delta i \sim 2.5$\degr. 
According to the analysis of \cite{Dai2018},
when the innermost planet has $a/$\rstar~$< 5$, the minimum mutual inclination with other planets in the system often reaches values up to $5$\degr-$10$\degr, with larger period ratios ($P_{\rm c}/P_{\rm b} > 5$-$6$) implying an higher mutual inclination. 
Considering the large period ratio of TOI-561 ($P_{\rm c}/P_{\rm b} \sim 24$) and the value of $a_{\rm b}/$\rstar~$=2.6$, the measured $\Delta i \sim 2.5$\degr\ in this case is much lower that the expected inclination dispersion of $6.7 \pm 0.7$\degr\ that \cite{Dai2018} inferred for systems with similar orbital configurations, indicating that the TOI-561 system probably evolved through a mechanism that did not excite the inclination of the innermost planet. \par

\begin{figure}
\centering
\includegraphics[width=\linewidth]{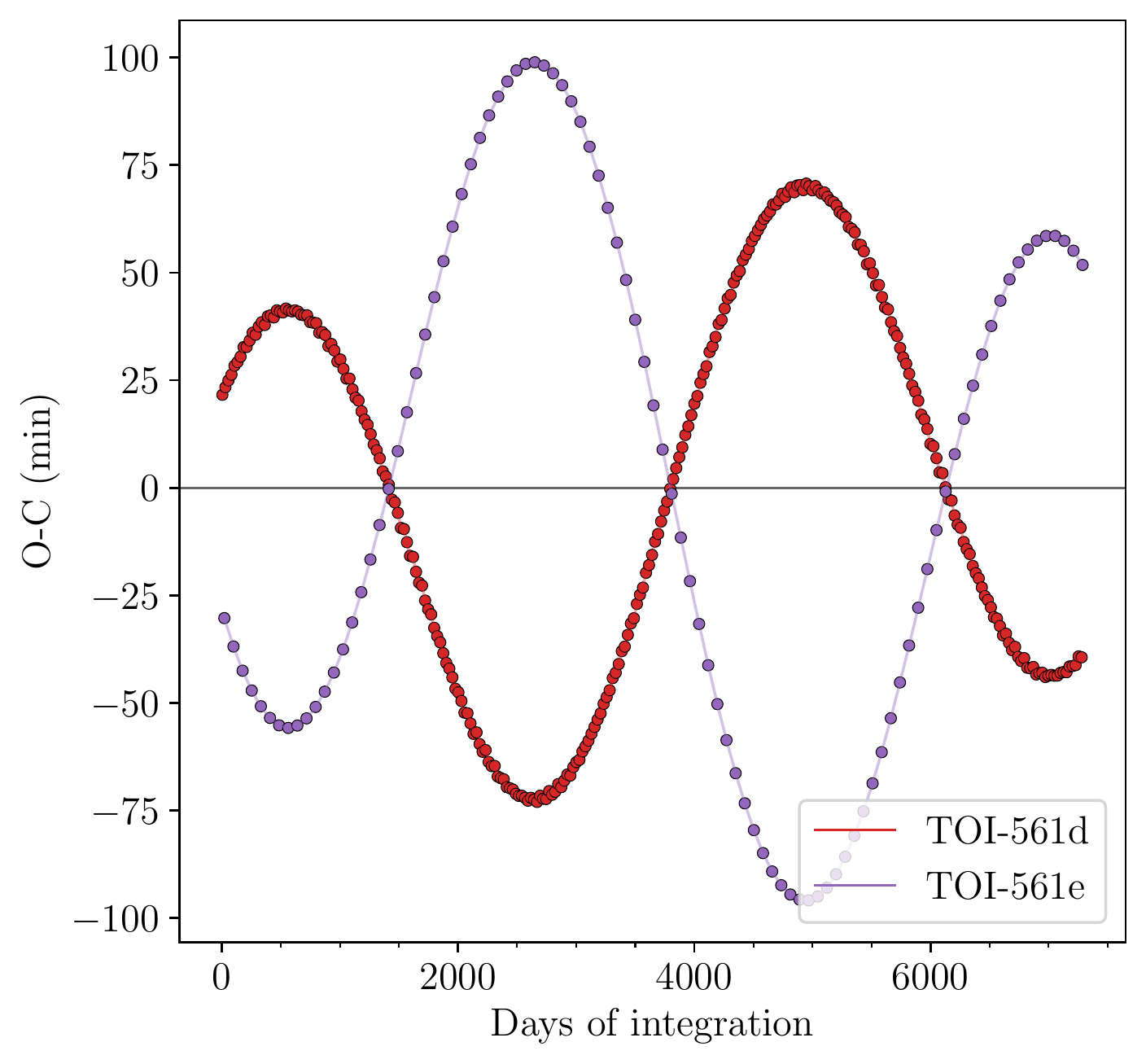}
  \caption{Predicted TTV signal of \starname\ d and e assuming our best-fitting model (see Table~\ref{table:joint_parameters}).
  The planets show a strong, anti-correlated signal.
  The signals of the USP planet ($<1$~sec) and of planet c ($<1$~min) are not reported.
  }
  \label{fig:TTV_simulation}
\end{figure}

We also performed a dynamical N-body simulation to check if significant TTVs are expected in the \starname\ system with our determined configuration.
In fact, the period ratio of \starname\ d and e indicates that
the planets are close to a 3:1 commensurability, hint of a second order mean motion resonance (MMR),
that may suggest the presence of a strong dynamical interaction between these planets. 
Starting from the initial configuration (as reported in Table~\ref{table:joint_parameters}),
we numerically integrated the orbits using the N-body integrator \texttt{ias15}
within the {\tt rebound} package \citep{rein2012}.
We assumed as reference time the $T_0$ of the USP planet
(see Table~\ref{table:joint_parameters}),
that roughly corresponds to the beginning of the {\it TESS} observations of TOI-561.
During the integration, we computed the transit times of each
planet following the procedure described in \cite{borsato2019}, and
we compared the inferred transit times with the linear ephemeris in order to obtain the TTV signal, 
reported as an observed-calculated diagram ($O-C$, \citealt{agol2018}) in Figure~\ref{fig:TTV_simulation}.
According to our simulation, \starname\ d and e display an anti-correlated TTV signal,
with a very long TTV period of $\sim 4850$~days ($\sim 13$~yr), and TTV amplitudes of $\sim 62$ minutes (planet d) and $\sim 84$ minutes (planet e), calculated computing the GLS periodogram of the simulated TTVs.
The anti-correlated signal 
demonstrates
that the two planets are expected to dynamically interact \citep{agol2018}.
In contrast, the predicted TTV amplitude of planet c is extremely low ($\sim 0.9$ min), being the planet far from 
any period commensurability,
as well as the USP planet, which has a
negligible TTV signal ($< 1$ sec).
With the solution for the planetary system we propose in this paper,
\starname\ 
is
a good target for a TTV follow-up, that will however require a very long
time
baseline in order to
tackle
the long-period TTV pattern. 
To better sample such a long-period TTV signal, it could be worth specifically re-observing the target when the deviations from the linear ephemeris are higher, \ie, during the periods corresponding to the $O-C$ peaks (or dips)
in Figure~\ref{fig:TTV_simulation}. 
According to our simulation,
the first peak (dip) corresponds to the period between March--December 2020,
while the second one will be between January--October 2026, \ie, corresponding to the time-spans between $\sim 400$--$700$ and $\sim 2500$--$3000$ days of integration in Figure~\ref{fig:TTV_simulation} respectively.
We remark that this calculation is performed assuming the $T_0$s inferred from single transit observations, thus implying a significant uncertainty in the TTV phase determination. 
Therefore, additional photometric observations
are necessary to refine the linear ephemeris of the planets,
and consequently also the prediction of the TTV phase. 

\subsection{Prospects for atmospheric characterization}\label{sec:atmo_char}

Given the interesting composition of the planets in the system, we checked if the \starname\ planets would be accessible targets for atmospheric characterisation through transmission spectroscopy, e.g. with the {\it James Webb Space Telescope} ({\it JWST}). 
For all the planets in the system, we calculated the Transmission Spectroscopy Metric (TSM, \citealt{kempton2018}), which predicts the expected transmission spectroscopy
SNR of a $10$-hour observing campaign with {\it JWST}/Near Infrared Imager and Slitless Spectrograph (NIRISS) under the assumptions of cloud-free atmospheres, the same atmospheric composition for all planets of a given type, and a fixed mass-radius relation. 
We obtained TSM values of $19$, $107$, $24$, and $14$ for planets b, c, d, and e, respectively. According to \citet{kempton2018}\footnote{The authors suggest to select planets with TSM $>12$ for \rplanet~$<1.5$ \mearth, TSM $>92$ for $1.5$~\rearth $<$ \rplanet~$<2.75$~\rearth, and TSM $>84$ for $2.75$~\rearth $<$ \rplanet~$<4$~\rearth.}, this classifies TOI-561 b and c as high-quality atmospheric characterisation targets among the {\it TESS} planetary candidates.
However, it should be noted that the TSM metric assumes rocky composition for planets with radius $< 1.5$~\rearth\, and according to our analysis TOI-561 b is not compatible with such a composition.
The same caveat holds for planet c, for which the assumptions under which the TSM is calculated may not be totally valid (e.g. the mass obtained from our analysis is not the same as if calculated with the \citet{chen&kipping2017} mass-radius relation, that is the relation assumed in \citet{kempton2018}, and that would imply a mass of $M_{\rm c} \simeq 8.7$~\mearth). 
Therefore, this estimate of the atmospheric characterisation feasibility should be used with caution, especially as the TSM metric has been conceived to prioritise targets for follow-up, and not to precisely determine the atmospheric transmission properties. 

\subsection{Summary and conclusions}\label{sec:summary}

According to our analysis, \starname\ hosts a nearly co-planar four-planet system, with an unusually low density USP super-Earth (planet b), a mini-Neptune (planet c) with a significant amount of volatiles surrounding a rocky core, and two mini-Neptunes, which are both consistent with a water-world scenario or with a rocky core surrounded by a gaseous envelope, and that are expected to show a strong, long-term TTV signal.
The multi-planetary nature of \starname\ offers a unique opportunity for comparative exoplanetology. \starname\ planets may be compared with the known population of multi-planet systems to understand their underlying distribution and occurrences, and to give insights on the formation and evolution processes of close-in planets, especially considering the intriguing architecture of the system, with the presence of a uncommonly low-density USP super-Earth and three mini-Neptunes on the opposite side of the radius valley. \par
Considering the few available data (\ie, $2$ transits for planet c, $1$ transit for planets d, e), additional observations are needed to unequivocally confirm our solution.
Further high-precision photometric (i.e. with {\it TESS}, that will re-observe \starname\ in sector $35$~--~February/March 2021, or with the {\it CHEOPS} satellite) and RVs observations will help improving the precision on the planets parameters, both allowing for the detection of eventual TTVs and increasing the time-span of the RV dataset, that could also unveil possible additional long-period companions.

\section*{Acknowledgements}
We thank the anonymous referee for the constructive comments and recommendations which helped improving the paper.\\
This paper includes data collected by the {\it TESS} mission,
which are publicly available from the Mikulski Archive for Space
Telescopes (MAST). Funding for the {\it TESS} mission is provided 
by the NASA Explorer Program. 
Resources supporting this work were provided by the NASA High-End Computing (HEC) Program through the NASA Advanced Supercomputing (NAS) Division at Ames Research Center for the production of the SPOC data products.
Based on observations made with the Italian 
Telescopio Nazionale Galileo (TNG) operated on the island of La Palma
by the Fundaci\'on Galileo Galilei of the INAF
at the Spanish Observatorio del Roque de los
Muchachos of the Instituto de Astrofisica de Canarias
(GTO program, and A40TAC\_23 program from INAF-TAC).
The HARPS-N project was funded by the Prodex Program of
the Swiss Space Office (SSO), the Harvard- University Origin
of Life Initiative (HUOLI), the Scottish Universities Physics
Alliance (SUPA), the University of Geneva, the Smithsonian
Astrophysical Observatory (SAO), and the Italian National
Astrophysical Institute (INAF), University of St. Andrews,
Queen's University Belfast and University of Edinburgh.
Parts of this work have been supported by the National Aeronautics and Space Administration under grant No. NNX17AB59G issued through the Exoplanets Research Program.
This research has
made use of the NASA Exoplanet Archive, which is
operated by the California Institute of Technology, 
under contract with the National Aeronautics and Space
Administration under the Exoplanet Exploration Program. 
This work has made use of data from the European Space Agency (ESA) mission
{\it Gaia} (\url{https://www.cosmos.esa.int/gaia}), processed by the {\it Gaia}
Data Processing and Analysis Consortium (DPAC,
\url{https://www.cosmos.esa.int/web/gaia/dpac/consortium}). 
Funding for the DPAC
has been provided by national institutions, in particular the institutions
participating in the {\it Gaia} Multilateral Agreement.
This publication makes use of data products from the Two
Micron All Sky Survey, which is a joint project of the
University of Massachusetts and the Infrared Processing and
Analysis Center/California Institute of Technology, funded by
the National Aeronautics and Space Administration and the
National Science Foundation. 
This work is made possible by a grant from the John Templeton Foundation. The opinions expressed in this publication are those of the authors and do not necessarily reflect the views of the John Templeton Foundation.\\
GL acknowledges support by CARIPARO Foundation, according to the agreement CARIPARO-Universit{\`a} degli Studi di Padova (Pratica n. 2018/0098), and scholarship support by the ``Soroptimist International d'Italia'' association (Cortina d'Ampezzo Club).
GLa, LBo, GPi, VN, GS, and IPa acknowledge the funding support from Italian Space Agency (ASI) regulated by ``Accordo ASI-INAF n. 2013-016-R.0 del 9 luglio 2013 e integrazione del 9 luglio 2015 CHEOPS Fasi A/B/C''.
DNa acknowledges the support from the French Centre National d'Etudes Spatiales (CNES).
AM acknowledges support from the senior Kavli Institute Fellowships.
ACC acknowledges support from STFC consolidated grant ST/R000824/1 and UK Space Agency grant ST/R003203/1.
ASB and MPi acknowledge financial contribution from the ASI-INAF agreement n.2018-16-HH.0.
XD is grateful to the Branco-Weiss Fellowship for
continuous support. This project has received funding from the
European Research Council (ERC) under the European Union's
Horizon 2020 research and innovation program (grant agreement
No. 851555). 
JNW thanks the Heising-Simons Foundation for support.
%%%%%%%%%%%%%%%%%%%%%%%%%%%%%%%%%%%%%%%%%%%%%%%%%%
\section*{Data Availability}
%The inclusion of a Data Availability Statement is a requirement for articles published in MNRAS. Data Availability Statements provide a standardised format for readers to understand the availability of data underlying the research results described in the article. The statement may refer to original data generated in the course of the study or to third-party data analysed in the article. The statement should describe and provide means of access, where possible, by linking to the data or providing the required accession numbers for the relevant databases or DOIs.

HARPS-N observations and data products are available through the Data \& Analysis Center for Exoplanets (DACE) at \url{https://dace.unige.ch/}. 
{\it TESS} data products can be accessed through the official NASA website \url{https://heasarc.gsfc.nasa.gov/docs/tess/data-access.html}.\\
All underlying data are available either in the appendix/online supporting material or will be available via VizieR at CDS.

%%%%%%%%%%%%%%%%%%%% REFERENCES %%%%%%%%%%%%%%%%%%

% The best way to enter references is to use BibTeX:

\bibliographystyle{mnras}
\bibliography{bibliography}

% Alternatively you could enter them by hand, like this:
% This method is tedious and prone to error if you have lots of references
%\begin{thebibliography}{99}
%\bibitem[\protect\citeauthoryear{Author}{2012}]{Author2012}
%Author A.~N., 2013, Journal of Improbable Astronomy, 1, 1
%\bibitem[\protect\citeauthoryear{Others}{2013}]{Others2013}
%Others S., 2012, Journal of Interesting Stuff, 17, 198
%\end{thebibliography}

%%%%%%%%%%%%%%%%%%%%%%%%%%%%%%%%%%%%%%%%%%%%%%%%%%

%%%%%%%%%%%%%%%%% APPENDICES %%%%%%%%%%%%%%%%%%%%%

\appendix

\section{Photometric analysis}\label{sec:appendix_transits}
\begin{figure*}
\centering
\includegraphics[width=\linewidth]{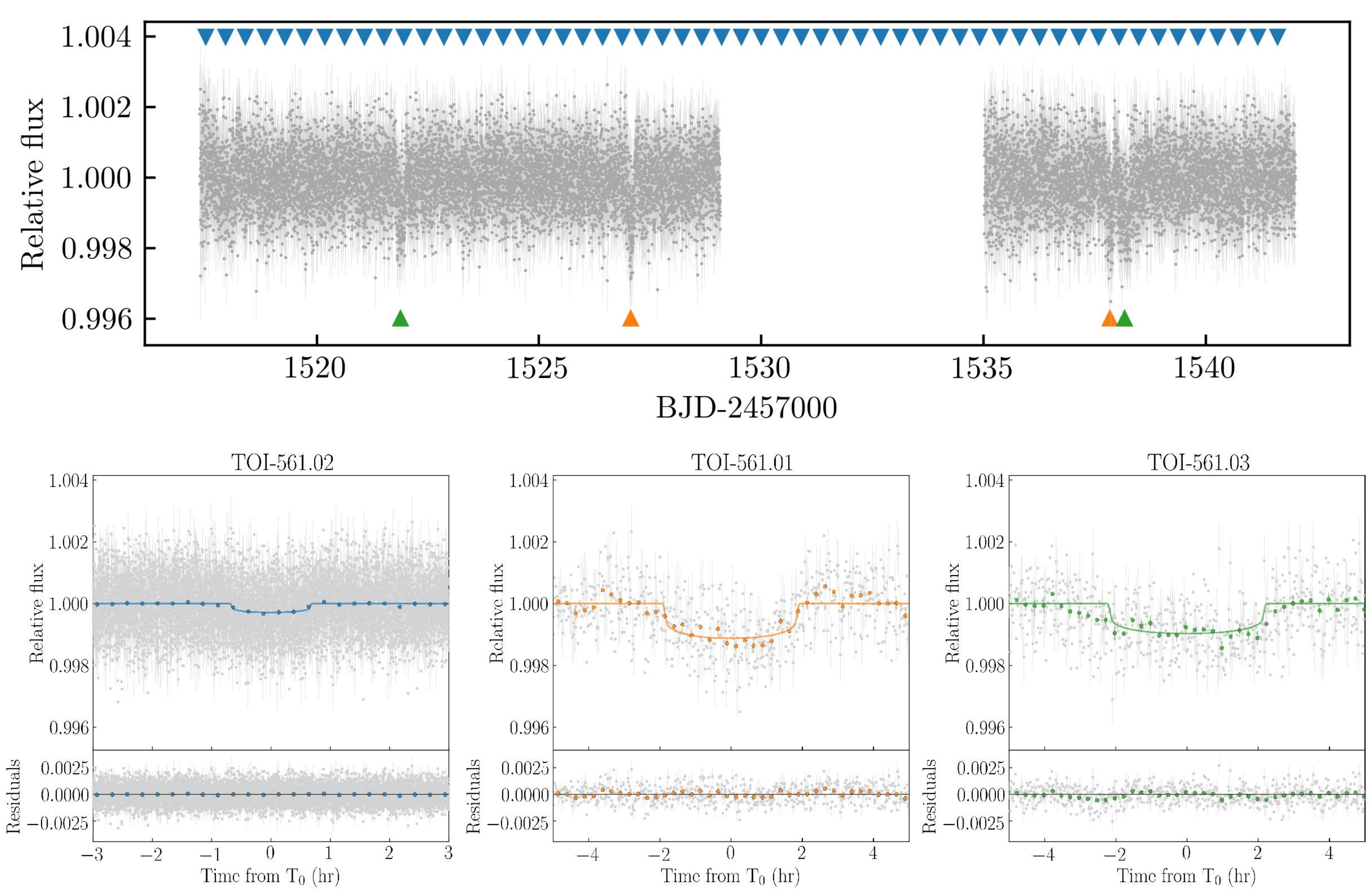}
  \caption{\emph{Top}: $2$-minute cadence flattened light curve of \starname.
  The transits of candidates TOI-561.02 ($P \sim0.45$~d), .01 ($P \sim10.8$~d), and .03 ($P \sim16.3$~d) are highlighted with blue, orange and green triangles, respectively. \emph{Bottom}: \starname\ phase-folded light curves over the best-fitting models (solid lines) for the three planets. The grey points are the {\it TESS} $2$-minute data, the coloured dots are the data points binned over $15$ minutes. The light curve residuals are shown in the bottom panel.
  Note the deviations from zero of the residuals in the ingress/egress phase for TOI-561.03.
    }
    \label{fig:photometry1}
\end{figure*}

We performed a preliminary light curve fit of the three candidate planets found by the SPOC pipeline and our independent TLS analysis,
that is TOI-561.01, .02, and .03 with periods of about $10.8$~d, $0.45$~d, and $16.3$~d, respectively.
We fit the transits using \pyorbit, as specified in Section~\ref{sec:data_analysis_tools}, but assuming circular orbits for all the candidate planets,
given the uncertainty associated with the eccentricity from the analysis of {\it TESS} data alone \citep{winn2010}.
 We ran the chains for $100\,000$ steps, discarding the first $20\,000$ as burn-in.
We list the obtained parameters in Table~\ref{table:lc_fit_params} and we show the best-fitting transit models in Figure \ref{fig:photometry1}.
In order to test whether our light curve flattening affected the inferred parameters of the planetary candidates, we also ran the \pyorbit\ fit on the original PDCSAP light curve. For all the candidates, the difference between the parameters of the two runs was lower than the error on the parameters themselves, indicating that the flattening did not significantly alter the results.\par
We stress that, at last, our global analysis disclaimed the presence of the planetary candidate TOI-561.03, linking the transits here associated with this candidate to single transits of two additional planets discovered in the system (see Section~\ref{sec:system_architecture}).

\begin{table*}
\caption{Planetary parameters of the three transiting candidates from the initial light curve fitting. }
\label{table:lc_fit_params}     
\centering
\begin{tabular}{l c c c}          
  \hline\hline                        
  Parameter & TOI-561.02 & TOI-561.01 & TOI-561.03 \\
  \hline
  $P$ (d) & $0.44656 \pm 0.00007$& $10.780 \pm 0.005$& $16.309_{-0.008}^{+0.010}$\\
  $T_0^{\rm a}$ (d) & $1517.4988 \pm 0.0019$ & $1527.060 \pm 0.004$  & $1521.884_{-0.006}^{+0.003}$ \\
  $a/$\rstar & $2.611 \pm 0.030$ & $21.81 \pm 0.25$ & $28.75 \pm 0.33$\\
  $a$ (AU) & $0.01055 \pm 0.00008$ &$0.0881 \pm 0.0007$ & $0.1161 \pm 0.0009$\\
  $R_\mathrm{p}/$\rstar & $0.01544 \pm 0.0007$ & $ 0.0308 \pm 0.0009$& $0.0285 \pm 0.0008$\\
  \rplanet\ (\rearth) & $1.46 \pm 0.06$& $2.91 \pm 0.10$ & $2.70 \pm 0.09$\\
  $b$ & $0.16_{-0.11}^{+0.14}$ & $0.17 \pm 0.12$ & $0.07_{-0.05}^{+0.07}$\\
  $i$ (deg) & $86.5_{-3.0}^{+2.7}$ & $89.54_{-0.33}^{+0.30}$ & $89.86_{-0.15}^{+0.10}$ \\
  $T_{14}^{\rm b}$ (hr) & $1.343_{-0.034}^{+0.022}$ & $3.82_{-0.10}^{+0.06}$& $4.44 \pm 0.06$ \\
 \hline
 \multicolumn{4}{c}{\emph{Common parameter}}\\
 \hline
  \rhostar\ (\rhosun) & $1.200 \pm 0.041$ & & \\
  $u_1$ & $0.381 \pm 0.047$ & & \\
  $u_2$ & $0.192 \pm 0.050$ & & \\
 \hline
\multicolumn{4}{l}{$^a$ BJD$_{\rm TDB}$-2457000.}\\
\multicolumn{4}{l}{\makecell[l]{$^b$ Transit duration is derived from the posterior distributions using the formulas \\ in \cite{seager2003}. }}
\end{tabular}
\end{table*}

\section{RV analysis}\label{sec:appendix_RV}
\subsection{Removal of anomalous points}\label{sec:appendix_removal}
Before proceeding with a 
detailed analysis,
we verified if any anomalous RV measurement was affecting our analysis. 
We followed a similar approach to that of \cite{Cloutier2019}, but slightly more sophisticated due to the presence of (possibly up to)
five planetary signals. 
Instead of analysing the power variation of the periodogram's peaks associated with the candidate planets while removing one point at the time, we decided to perform a full RV fit with the methodology described in Section~\ref{sec:data_analysis_tools}, 
and to compare the resulting RV semi-amplitudes with those derived using the full dataset.
To reduce computational time, we decided to remove from the dataset $5$ consecutive observations at once (\ie, performing $17$ iterations rather than $82$), and then performed the leave-one-out cross-validation on those subsets showing deviating RV semi-amplitudes in order to identify the anomalous RV measurement. With this approach, we found out that a total of $5$ RV measurements, with associated errors greater than $2.5$~\ms\ and \snr~$< 35$ were systematically producing a decrease in the semi-amplitude of candidates .01 and .02 by $\approx 0.1-0.2$ \ms, and we therefore removed these points from our dataset in order to improve the accuracy of our results, even if the total variation in RV semi-amplitude was within the error bars.
We note that these observations are clearly outliers at more than $2 \sigma$ in both the \snr~of the spectra and the RV error distributions (see Section \ref{sec:harpsn_RV}), which is simply the consequence of having been gathered in sub-optimal weather conditions.
A much simpler sigma-clipping selection would have led to the exclusion of the same data points. 
The complex approach we employed in this work can thus be avoided in future analysis involving HARPS-N data. 

\subsection{RV modelling and injection/retrival tests}\label{sec:appendix_RV_model}
Given the results of the frequency analysis in Section~\ref{sec:RV_signals}, we performed a \texttt{PyDE}+\texttt{emcee} RV fit with \pyorbit,
following the methodology 
as described in Section~\ref{sec:data_analysis_tools}, and
assuming the model suggested by the Bayesian evidence computed in Section~\ref{sec:RV_signals} (see Table~\ref{table:logZ}), \ie\ a model with the three transiting candidates plus two additional ones. 
We performed two independent fits, constraining the period of the outer signal to be 
shorter or longer
than $100$ days, in order to disentangle the $78$ periodicity from its alias at $180$ respectively. 
We ran the chains for $150\,000$ steps, discarding the first $50\,000$ as burn-in.
The results of this analysis are reported in Tables~\ref{table:RV_fit_params_78d} and \ref{table:RV_fit_params_180d}. 

\begin{table*}
\caption{Best-fitting parameters from the five-planet RV fit, assuming period boundaries of $2$-$100$ days for the outermost planet.}
\label{table:RV_fit_params_78d}     
\centering
\begin{tabular}{l c c c c c}         
  \hline\hline                        
  Parameter & TOI-561.02 & TOI-561.01 & TOI-561.03 & TOI-561.04 & TOI-561.05\\
  \hline
  $P$ (d) & $0.44658 \pm 0.00001$& $10.778 \pm 0.004$& $16.294 \pm 0.008$ & $25.64_{-0.18}^{+0.21}$ & $77.9 \pm 1.9$\\
  $T_0^{\rm a}$ (d) & $1517.4983 \pm 0.0008$ & $1527.061 \pm 0.003$  & $1521.883 \pm 0.004$ & $1521_{-5}^{+3}$ & $1535_{-10}^{+9}$\\
  $e$ & $0$ (fixed) & $0.069_{-0.048}^{+0.068}$& $0.069_{-0.048}^{+0.074}$ & $0.073_{-0.051}^{+0.078}$ & $0.061_{-0.043}^{+0.068}$ \\
  $\omega$ (deg) & $90$ (fixed) & $178 \pm 75$ & $235_{-100}^{+135}$ & $275_{-80}^{+60}$ & $100_{-113}^{+93}$\\
  $K$ (\ms) &$1.41 \pm 0.33$ & $1.73 \pm 0.36$& $<0.37$ & $3.12 \pm 0.36$ & $2.78 \pm 0.44$\\
  \mplanet\ (\mearth) &$1.43 \pm 0.33$ & $5.1 \pm 1.0$& $<1.27$ &  $12.2 \pm 1.4$ & $15.7 \pm 2.5$\\
 \hline
 \multicolumn{6}{c}{\emph{Common parameter}}\\
 \hline
  $\sigma_{\rm jitter}^{\rm b}$ (\ms) & $1.32 \pm 0.23$ & & & & \\
  $\gamma^{\rm c}$ (\ms) & $79702.58 \pm 0.30$ & & & & \\
 \hline
\multicolumn{6}{l}{$^a$ BJD$_{\rm TDB}$-2457000.} \\
\multicolumn{6}{l}{$^b$ Uncorrelated jitter term.}\\
\multicolumn{6}{l}{$^c$ RV offset.}\\
\end{tabular}
\end{table*}

\begin{table*}
\caption{Best-fitting parameters from the five-planet RV fit, assuming period boundaries of $100$-$200$ days for the outermost planet.}
\label{table:RV_fit_params_180d}      
\centering
\begin{tabular}{l c c c c c}         
  \hline\hline                        
  Parameter & TOI-561.02 & TOI-561.01 & TOI-561.03 & TOI-561.04 & TOI-561.05\\
  \hline
  $P$ (d) & $0.44658 \pm 0.00001$& $10.779 \pm 0.004$& $16.294 \pm 0.007$ & $25.82 \pm 0.19$ & $179.5_{-7.4}^{+8.3}$\\
  $T_0^{\rm a}$ (d) & $1517.4983 \pm 0.0009$ & $1527.061 \pm 0.003$  & $1521.883 \pm 0.004$ & $1518 \pm 3$ & $1633_{-15}^{+13}$\\
  $e$ & $0$ (fixed) & $0.067_{-0.047}^{+0.072}$& $0.064_{-0.045}^{+0.070}$ & $0.072_{-0.051}^{+0.071}$ & $0.058_{-0.041}^{+0.064}$ \\
  $\omega$ (deg) & $90$ (fixed) & $148_{-107}^{+118}$ & $189_{-127}^{+118}$ & $287_{-73}^{+67}$ & $128_{-113}^{+98}$\\
  $K$ (\ms) &$1.57 \pm 0.32$ & $0.69_{-0.46}^{+0.41}$& $<0.54$ & $3.10 \pm 0.36$ & $3.17 \pm 0.49$\\
  \mplanet\ (\mearth) &$1.59 \pm 0.33$ & $2.01_{-1.35}^{+1.20}$& $<1.91$ &  $12.1 \pm 1.4$ & $23.7 \pm 3.7$\\
 \hline
 \multicolumn{6}{c}{\emph{Common parameter}}\\
 \hline
  $\sigma_{\rm jitter}^{\rm b}$ (\ms) & $1.34 \pm 0.23$ & & & & \\
  $\gamma^{\rm c}$ (\ms) & $79703.86 \pm 0.25$ & & & & \\
 \hline
 \multicolumn{6}{l}{$^a$ BJD$_{\rm TDB}$-2457000.} \\
\multicolumn{6}{l}{$^b$ Uncorrelated jitter term.}\\
\multicolumn{6}{l}{$^c$ RV offset.}\\
\end{tabular}
\end{table*}

In all our RV fits, regardless of the assumed period of the outermost planet, 
TOI-561.03 (\ie, the candidate with period of $\sim 16.3$~d) remains undetected with an upper limit of $K \lesssim 0.5$~\ms , corresponding to a rather nonphysical mass of $\lesssim 2$~\mearth\ (at $1\sigma$) for a planet with \rplanet~$ \simeq 2.7 $~\rearth.
We thus performed a series of injection/retrieval simulations in order to assess the influence of the observational sampling and of the precision in the mass measurements of the other planets. 
In a first run, the synthetic datasets were simulated
by assuming the orbital parameters as previously determined in the RV fits for the candidate
planets .01, .02,
and the non-transiting candidates, 
while the RV semi-amplitude of the candidate planet at $16$~d was varied between $0.0$~\ms\ and $1.5$~\ms\ in steps of $0.5$~\ms. 
For computational reasons, we performed this analysis only with the $78$-d solution for the outer planet.
We projected the model onto the real epochs of observation and then we added a Gaussian noise corresponding to the measured error plus an RV jitter of $1.0$~\ms\ added in quadrature, while preserving the original value in the analysis. We built $50$ different noise realisations and analysed each of 
them
with the same methodology as before, \ie , {\tt PyDE}+{\tt emcee} through \pyorbit, but for a shorter chain length\footnote{$10\,000$ steps after convergence, reached at approximately $15\,000$ steps.} to reduce computing time. The posteriors of each parameter were then obtained by putting together the individual posterior distributions from each noise realisation.
We finally repeated the same analysis but varying the RV semi-amplitude 
of the candidate
planet
.01, \ie , the closest signal in frequency space and the one with the most uncertain RV semi-amplitude measurement other than the USP candidate, by $\pm 0.5$~\ms\ with respect to the value of $1.7$~\ms\ used in the previous analysis. 

\section*{Affiliations}
\noindent
{\it
$^{1}$Department of Physics and Astronomy, Universit{\`a} degli Studi di Padova, Vicolo dell'Osservatorio 3, I-35122, Padova, Italy\\
$^{2}$INAF - Osservatorio Astronomico di Padova, Vicolo dell'Osservatorio 5, I-35122, Padova, Italy\\
$^{3}$Aix-Marseille Universit{\'e}, CNRS, CNES, LAM, F-13013 Marseille, France\\
$^{4}$Astrophysics Group, Cavendish Laboratory, University of Cambridge, J.J. Thomson Avenue, Cambridge CB3 0HE, UK\\
$^{5}$Kavli Institute for Cosmology, University of Cambridge, Madingley Road, Cambridge CB3 0HA, UK\\
$^{6}$Observatoire Astronomique de l'Universit{\'e} de Gen{\`e}ve, 51 Chemin des Maillettes, 1290 Versoix, Switzerland\\
$^{7}$Centre for Exoplanet Science, SUPA, School of Physics and Astronomy, University of St Andrews, St Andrews KY16 9SS, UK\\
$^{8}$Fundaci\'on Galileo Galilei-INAF, Rambla J. A. F. Perez, 7, E-38712, S.C. Tenerife, Spain\\
$^{9}$INAF-Osservatorio Astronomico di Brera, via E. Bianchi 46, 23807 Merate (LC), Italy\\
$^{10}$DTU Space, National Space Institute, Technical University of Denmark, Elektrovej 328, DK-2800 Kgs. Lyngby, Denmark\\
$^{11}$Center for Astrophysics | Harvard \& Smithsonian, 60 Garden Street, Cambridge, MA, 02138, USA\\
$^{12}$INAF - Osservatorio Astrofisico di Torino, Via Osservatorio 20, I-10025, Pino Torinese, Italy\\
$^{13}$NASA Ames Research Center, Moffett Field, CA, 94035, USA\\
$^{14}$SUPA, Institute for Astronomy, University of Edinburgh, Blackford Hill, Edinburgh, EH9 3HJ, Scotland, UK\\
$^{15}$Centre for Exoplanet Science, University of Edinburgh, Edinburgh, EH93FD, UK\\
$^{16}$Department of Astrophysical Sciences, Princeton University, 4 Ivy Lane, Princeton, NJ 08544, USA\\
$^{17}$INAF - Osservatorio Astronomico di Roma, Via Frascati 33, 00078, Monte Porzio Catone, Italy\\
$^{18}$INAF - Osservatorio Astronomico di Cagliari, via della Scienza 5, 09047, Selargius, Italy\\
$^{19}$ INAF - Osservatorio Astronomico di Palermo, Piazza del Parlamento 1, I-90134 Palermo, Italy\\
$^{20}$INAF - Osservatorio Astrofisico di Catania, Via S. Sofia 78, I-95123, Catania, Italy\\
$^{21}$Department of Earth, Atmospheric and Planetary Sciences, and Kavli Institute for Astrophysics and Space Research, Massachusetts Institute of Technology, Cambridge, MA 02139, USA\\
$^{22}$Astrophysics Research Centre, School of Mathematics and Physics, Queen's University Belfast, Belfast, BT7 1NN, UK
}

% Don't change these lines
\bsp	% typesetting comment
\label{lastpage}
\end{document}